\documentclass[12pt,aps,prd,showkeys,showpacs,groupedaddress]{revtex4}
\usepackage{amsfonts}
\usepackage{amsmath}
\usepackage{amssymb}
\usepackage{bm}
\expandafter\ifx\csname package@font\endcsname\relax\else
\expandafter\expandafter
\expandafter\usepackage
\expandafter\expandafter
\expandafter{\csname package@font\endcsname}%
\fi
\newcommand{\il}{\mathcal{I}_{A_l}}

\newcommand{\ilrrp}{\ffrac{{\mathcal{I}}^{(p)}_{A_l}(t-r)}r}

\newcommand{\eislrrp}{\ffrac{\epsilon_{iba}{\mathcal{S}}^{(p)}_{bA_{l-1}}(t-r)}r}

\newcommand{\eisbr}{\ffrac{\epsilon_{iba}\mathcal{S}_b}r}

\newcommand{\Mc}{\mathcal M}
\newcommand{\Ic}{\mathcal I}
\newcommand{\Sc}{\mathcal S}
\newcommand{\ffrac}[2]{\frac{{\displaystyle #1}}{{\displaystyle #2}}}

\newcommand{\dksi}{\hat\partial}
\newcommand{\dtau}{\hat\partial_\tau}
\newcommand{\dtautau}{\hat\partial_{\tau\tau}}
\newcommand{\dt}{\hat\partial_{t^*}}

\newcommand{\phim}{\varphi_{\scriptscriptstyle(M)}}
\newcommand{\phis}{\varphi_{\scriptscriptstyle(S)}}
\newcommand{\chim}{\chi_{\scriptscriptstyle(M)}}
\newcommand{\chis}{\chi_{\scriptscriptstyle(S)}}
\newcommand{\fim}{\phi_{\scriptscriptstyle(M)}}
\newcommand{\fis}{\phi_{\scriptscriptstyle(S)}}
\newcommand{\fig}{\phi_{\scriptscriptstyle(G)}}

\newcommand{\DXGi}{\DXii\limits_{\scriptscriptstyle (G)}{}\left(\tau,{\bm \xi}\right)}

\newcommand{\XG}{\Xii\limits_{\scriptscriptstyle (G)}{}\left(\tau,{\bm \xi}\right)}
\newcommand{\XGo}{\Xii\limits_{\scriptscriptstyle (G)}{}\left(\tau_0,{\bm \xi}\right)}
\newcommand{\DXiim}{\DXii\limits_{\scriptscriptstyle (M)}{}\left(\tau,{\bm \xi}\right)}
\newcommand{\DXiis}{\DXii\limits_{\scriptscriptstyle (S)}{}\left(\tau,{\bm \xi}\right)}
\newcommand{\Xiim}{\Xii\limits_{\scriptscriptstyle (M)}{}\left(\tau,{\bm \xi}\right)}
\newcommand{\Xiis}{\Xii\limits_{\scriptscriptstyle (S)}{}\left(\tau,{\bm \xi}\right)}

\newcommand{\Xiimo}{\Xii\limits_{\scriptscriptstyle (M)}{}\left(\tau_0,{\bm \xi}\right)}
\newcommand{\Xiiso}{\Xii\limits_{\scriptscriptstyle (S)}{}\left(\tau_0,{\bm \xi}\right)}

\newcommand{\Xijm}{\Xij\limits_{\scriptscriptstyle (M)}{}\left(\tau,{\bm \xi}\right)}
\newcommand{\Xijs}{\Xij\limits_{\scriptscriptstyle (S)}{}\left(\tau,{\bm \xi}\right)}

\newcommand{\DelG}{\Del\limits_{\scriptscriptstyle (G)}{}\left(\tau,\tau_0\right)}
\newcommand{\Delm}{\Del\limits_{\scriptscriptstyle (M)}{}\left(\tau,\tau_0\right)}
\newcommand{\Dels}{\Del\limits_{\scriptscriptstyle (S)}{}\left(\tau,\tau_0\right)}

\def\h{\operatornamewithlimits{h}}
\def\q{\operatornamewithlimits{q}}
\def\D^i{\operatornamewithlimits{D^{\it i}}}
\def\Del{\operatornamewithlimits{\Delta}}
\def\DXii{\operatornamewithlimits{\dot\Xi^{\it i}}}
\def\Xii{\operatornamewithlimits{\Xi^{\it i}}}
\def\DXij{\operatornamewithlimits{\dot\Xi^{\it j}}}
\def\Xij{\operatornamewithlimits{\Xi^{\it j}}}
\makeatletter \@addtoreset{equation}{section} \makeatother
\def\Quadrat#1#2{{\vcenter{\hrule height #2
  \hbox{\vrule width #2 height #1 \kern#1
    \vrule width #2}
  \hrule height #2}}}
\def\dAl{\mathop{\kern 1pt\hbox{$\Quadrat{8pt}{0.4pt}$} \kern1pt}}
\makeatletter
\@addtoreset{equation}{section}
\makeatother

\begin{document}
\title[Propagation of Light in the Field of Gravitational Multipoles]{Propagation of Light in the Field of Stationary and Radiative Gravitational Multipoles}
\author{Sergei Kopeikin\footnote{To
whom correspondence should be addressed (kopeikins@missouri.edu)}}\email{kopeikins@missouri.edu} 
\author{Pavel Korobkov} \email{PavelKorobkov@mizzou.edu}
\affiliation{Department of Physics \& Astronomy,
University of Missouri-Columbia, Columbia, MO 65211, USA} 
\author{Alexander Polnarev}
\email{A.G.Polnarev@qmul.ac.uk}
\affiliation{Queen Mary University of London, London E14NS, UK}
\date{\today}
\begin{abstract}
\noindent
Extremely high precision of near-future radio/optical interferometric observatories like SKA, Gaia, SIM
and the unparalleled sensitivity of LIGO/LISA gravitational-wave detectors demands more deep theoretical treatment of relativistic effects in the
propagation of electromagnetic signals through variable gravitational fields of the solar system, oscillating and precessing neutron stars, coalescing binary systems, exploding supernova,
and colliding galaxies. Especially important for future gravitational-wave observatories is the problem of propagation
of light rays in the field of multipolar gravitational waves emitted by a localized source
of gravitational radiation. 
Present paper suggests physically-adequate and consistent solution of this problem in the first post-Minkowskian approximation of General Relativity which accounts for  
all time-dependent multipole moments of an isolated astronomical system. 
We derive equations of propagation of electromagnetic wave in the retarded 
gravitational field of the localized source emitting 
gravitational waves of arbitrary multipolarity and integrate them analytically in closed form. We also prove that the leading terms in the observable relativistic effects (time delay, deflection angle, and rotation of the plane of polarization of light) depend on the instantaneous value of the multipole moments of the isolated system and its time derivatives taken at the retarded instant of time but not on their integrated values.   
The influence 
of the multipolar gravitational field of the isolated system on the light propagation is examined for a general case when light propagates not only through the wave zone of the system but also through its intermediate and near zones. The gauge freedom of our formalism is carefully studied and all gauge-dependent terms are singled out and separated from 
observable quantities. 
We also present a thorough-going
analytical treatment of time delay, light-ray bending and polarization of light in 
the case of large
impact parameter corresponding to the approximation of a plane gravitational wave of arbitrary multipolarity. This exploration
essentially extends previous results regarding propagation of
light rays in the quadrupolar field of a 
plane monochromatic gravitational wave. Explicit 
expressions for time delay, deflection angle and rotation of the plane of polarization of light are
obtained in terms of the transverse-traceless (TT) part of the space-space 
components of the metric tensor. We also discuss the relevance 
of the developed formalism for detection of relativistic effects of multipolar gravitational fields by existing and near-future astronomical techniques as 
well as the gravitational wave detectors.
\end{abstract}
\keywords{gravitation -- relativity -- experimental gravity -- gravitational waves}
\pacs{04.30.-w, 04.80.-y, 04.80.Nn, 95.55.Ym, 95.85.Sz }
\maketitle
\pagebreak
\tableofcontents 
\pagebreak
\section{Introduction}
\subsection{Statement of the Problem and Relation to Previous Work.}
We consider propagation of electromagnetic signals through the
time-dependent gravitational field of an isolated gravitating
system emitting gravitational waves such as, for example, a binary
system or oscillating neutron star. Gravitational field is assumed to be weak everywhere along the light
ray and a linear (post-Minkowskian) approximation of General Relativity is used. We do
not restrict our consideration to a specific system, but rather,
following \cite{KSGU99}, develop a formalism which can be
applied to an arbitrary self-gravitating source that is completely characterized by its time-dependent 
multipole moments of arbitrary order. We assume the wavelength of
light to be much smaller than the wavelength of the gravitational
waves which allows us to work in the geometrical optics approximation. The
relative distances between and positions of the source of light, the isolated system,
and the observer are not restricted as well, which makes our
mathematical formalism quite general and applicable for most practical
situations. With the assumptions made, we show that the equations
of light propagation, which in geometrical optics are the
equations of null geodesics, can be analytically integrated giving
linearized metric perturbations to the unperturbed flat space-time 
trajectory. We also use the formalism developed in \cite{AnBr74}
and \cite{KoMa02} to consider the propagation of the polarization
of light along the light ray and, in particular, the effect of
gravity-induced rotation of the polarization plane of an
electromagnetic wave (the Skrotskii effect).

This work is essentially a generalization of \cite{KSGU99} for the
case when the isolated system possesses multipole moments of
arbitrary order and type. In \cite{KSGU99} the same problem is considered
in spin-dipole and mass-quadrupole approximation only. We also generalize
the results of \cite{KoMa02} for Skrotskii effect on the case of
source possessing higher order multipole moments. The results of
\cite{KoMa02} for isolated system emitting gravitational waves
were obtained again only in spin-dipole mass-quadrupole approximation.
Formalism of the present paper allows us to extend results of  
exploration of propagation of electromagnetic waves in random field of plane gravitational waves \cite{BKNP1,BKNP2} to more general class of an isolated systems emitting non-planar gravitational waves anisotropically. 
Additional mathematical extention of the present work with more thorough historical review isgiven in the next publication \cite{KoKo05}.

\subsection{Notations and Conventions}
Notations in this paper coincide with those of \cite{KoKo05}.
Metric tensor on the space-time manifold is denoted by
$g_{\alpha\beta}$
 and its perturbation $h_{\alpha\beta}=g_{\alpha\beta}-\eta_{\alpha\beta}$.
 The determinant of the metric tensor is negative and is denoted as $g={\rm det}||g_{\alpha\beta}||$.
 The four-dimensional fully antisymmetric Levi-Civita symbol $\epsilon_{\alpha\beta\gamma\delta}$ is
 defined in accordance with the convention $\epsilon_{0123}=+1$.
 In the present paper we use a geometrodynamic system of
units \cite {LaLi62,MTW73} such that $c=G=1$ where c is the
fundamental speed and G is the universal gravitational constant.
Space-time is assumed to be asymptotically-flat and covered by a
single coordinate system $ (x^0,x^1,x^2,x^3)=(t,x,y,z)\;, $ where
$t$ and $(x,y,z)$ are time and space coordinates respectively.
This coordinate system is reduced at infinity to the Minkowskian
coordinates. We shall also use the spherical coordinates $
(r,\theta,\phi) $ related to $(x,y,z)$ by the standard
transformation
\begin{equation}
x=r\sin\theta\cos\phi\;,\quad y=r\sin\theta\sin\phi\;,\quad
z=r\cos\theta\;.
\end{equation}
Greek (spacetime) indices range from 0 to 3, and Latin (space)
indices run from 1 to 3. If not specifically stated the opposite,
the indices are raised and lowered by means of the Minkowski
metric $ \eta_{\alpha\beta}\equiv{\rm diag}(-1,1,1,1). $ Regarding
this rule the following conventions for coordinates hold:
$x^i=x_i$ and $ x^0=-x_0$. We also adopt notations,
$\delta_{ij}\equiv{\rm diag}(1,1,1)$, for the Kroneker symbol (a
unit matrix), and, $\epsilon_{ijk}$, for the fully antisymmetric
3-dimensional symbol of Levi-Civita with convention
$\epsilon_{123}=+1$.

Repeated indices are summed over in accordance with the Einstein's
rule \cite{LaLi62,MTW73}. In the linearized approximation of
general relativity used in this work there is no difference
between spatial vectors and co-vectors or between upper
and lower space indices. Therefore, for a dot product of two space
vectors we have
\begin{equation}
A^iB_i=A_iB_i\equiv A_1B_1+A_2B_2+A_3B_3\;.
\end{equation}
In what follows, we shall commonly use spatial multi-index
notations for Cartesian three-dimensional tensors \cite{Tho80},
that is
\begin{equation}
 \il\equiv \mathcal{I}_{a_1\ldots a_l}.
\end{equation}
Tensor product of $l$ identical spatial vectors $k^i$ will be
denoted as a three-dimensional tensor having $l$ indices
\begin{equation}
 k_{a_1\ldots}k_{a_l}\equiv k_{a_1\ldots a_l}\;. \end{equation}
Full symmetrization with respect to a group of spatial indices of
a Cartesian tensor will be distinguished with round brackets put
around the indices
\begin{equation}
Q_{(a_1\ldots a_l)}\equiv \frac{1}{l!}\sum_\sigma
Q_{\sigma(1)\ldots\sigma(l)}\;,
\end{equation}
where $\sigma$ is the set of all permutations of $(1,2,...,l)$
which makes $Q_{a_1\ldots a_l}$ fully symmetrical in $a_1\ldots
a_l$.

It is convenient to introduce a special notation for symmetric
trace-free (STF) Cartesian tensors by making use of angular
brackets around STF indices. The explicit expression of the STF
part of a tensor $Q_{a_1\ldots a_l}$ is \cite{Tho80}
\begin{equation}
Q_{<a_1\ldots
a_l>}\equiv\sum_{k=0}^{[l/2]}a^l_k\delta_{(a_1a_2}\cdot\cdot\cdot\delta_{a_{2k-1}a_{2k}}S_{a_{2k+1}\ldots
a_l)b_1b_1\ldots b_kb_k}\;,
\end{equation}
where $[l/2]$ is the integer part of $l/2$,
\begin{equation}S_{a_{1}\ldots a_l}\equiv Q_{(a_1\ldots a_l)}
\end{equation} and numerical coefficients
\begin{equation}
a^l_k=\frac{(-1)^k}{(2k)!!}\frac{l!}{(2l-1)!!}\frac{(2l-2k-1)!!}{(l-2k)!}\;.
\end{equation}
We also assume that for any integer $l\ge 0$
\begin{equation}
 l!\equiv l(l-1)\ldots 2\cdot 1\;,\qquad 0!\equiv 1\;,
\end{equation}
and
\begin{equation}
 l!!\equiv l(l-2)(l-4)\ldots (2\mbox{ or} 1)\;,\qquad 0!!\equiv 1\;.
\end{equation}
One has, for example,
\begin{equation} T_{<abc>}\equiv T_{(abc)}
                    -\frac{1}5\delta_{ab}T_{(cjj)}
                      -\frac{1}5\delta_{bc}T_{(ajj)}
                        -\frac{1}5\delta_{ac}T_{(bjj)}\;.
\end{equation}
Cartesian tensors of the mass-type $\mathcal {I}_{<A_l>}$ and
spin-type multipoles $\mathcal {S}_{<A_l>}$ entirely describing
gravitational field outside of an isolated astronomical system
 are always STF objects that can be checked by inspection of the definition following from the multipolar
 decomposition of the metric tensor perturbation $h_{\alpha\beta}$ \cite{Tho80}. For this reason,
  to avoid appearance of too complicated index notations we shall omit in the following text the angular
   brackets around indices of these (and only these) tensors, that is we adopt: $\il\equiv\mathcal {I}_{<A_l>}$
    and $\mathcal {S}_{A_l}\equiv\mathcal {S}_{<A_l>}$.

We shall also use transverse (T) and transverse-traceless (TT)
Cartesian tensors in our calculations \cite{MTW73,Tho80}. These
objects are defined by making use of the operator of projection $
P_{jk}\equiv \delta_{jk}-k_{jk} $ onto the plane orthogonal to a
unit vector $k_j$. Thus, one has \cite{Tho80}
\begin{eqnarray}
Q_{a_1\ldots a_l}^{\rm T}&\equiv&
P_{a_1b_1}P_{a_2b_2}...P_{a_lb_l}Q_{b_1\ldots b_l}\;,
\\
Q_{a_1\ldots a_l}^{\rm TT}&\equiv&\sum_{k=0}^{[l/2]}b^l_k
P_{(a_1a_2}\cdot\cdot\cdot P_{a_{2k-1}a_{2k}}W_{a_{2k+1}\ldots
a_l)b_1b_1\ldots b_kb_k}\;,
\end{eqnarray}
where again $[l/2]$ is the integer part of $l/2$,
\begin{equation}W_{a_{1}\ldots a_l}\equiv Q_{(a_1\ldots a_l)}^{\rm
T} \end{equation} and numerical coefficients
\begin{equation}
b^l_k=\frac{(-1)^k}{4^{k}}\frac{l(l-k-1)!!}{k!(l-2k)!}\;.
\end{equation}
For instance,
\begin{equation}
Q_{ab}^{\rm TT}\equiv
\frac{1}{2}\left(P_{ai}P_{bj}Q_{ij}+P_{bi}P_{aj}Q_{ij}\right)-\frac{1}2P_{ab}\left(P_{jk}Q_{jk}\right)\;.
\end{equation}

Polynomial coefficients will be used in some of our equations and
they are defined by
\cite{GRyzh}
\begin{equation}\label{pco}
C_l(p_1,\ldots ,p_n)\equiv \frac{l!}{p_1!\ldots p_n!}\;,
\end{equation}
where $l$ and $p_i$ are positive integers such that
$\sum_{i=1}^{n}p_i=l$. We introduce a Heaviside unit step function
$H(p-q)$ such that on the set of whole numbers
\begin{equation}
\label{H}
  H(p-q)=\begin{cases}0,& \text{if $p\leq q$,}\\
                  1,&\text{if $p > q$}.
                 \end{cases}
\end{equation}
For any differentiable function $f=f(t,{\bm x})$ one uses
notations: $f_{,0}=\partial f/\partial t$ and $f_{,i}=\partial
f/\partial x^i$ for its partial derivatives. In general, comma
standing after the function denotes a partial derivative with
respect to a corresponding coordinate: $f,_\alpha\equiv \partial
f(x)/\partial x^\alpha $. Overdot denotes a total derivative with
respect to time $ \dot f\equiv df/dt =\partial f/\partial t+\dot
x^i\partial f/\partial x^i$ usually taken in this paper along the
light ray trajectory ${\bm x}(t)\equiv (x^i)$. Sometimes the partial
derivatives with respect to space coordinate $x^i$ will be also
denoted as $\partial_i\equiv \partial /\partial x^i$, and the
partial time derivative will be denoted as $\partial_t\equiv
\partial /\partial t$. Covariant derivative with respect to the
coordinate $x^\alpha$ will be denoted as $\nabla_\alpha$.

We shall use special notations for integrals with respect to time
and for those taken along the light-ray trajectory. Specifically,
the time integrals from a function $F(t, {\bm x})$, where ${\bm
x}$ is not specified, are denoted as
\begin{equation}
\label{i1}
         F^{(-1)}(t, {\bm x})\equiv \int\limits_{-\infty}^t
                                                   F(\tau, {\bm x}) d\tau\;,
             \qquad\qquad
         F^{(-2)}(t, {\bm x})\equiv \int\limits_{-\infty}^t
                             F^{(-1)} (\tau, {\bm x}) d\tau\;.
\end{equation}
Time integrals from the function $F(t,{\bm x})$ taken along the
light ray so that spatial coordinate ${\bm x}$ is a function of
time ${\bm x}\equiv{\bm x}(t)$, are denoted as
\begin{equation}
\label{i2}
          F^{[-1]}(t,{\bm x})\equiv
                    \int\limits_{-\infty}^{t}
                       F(\tau,{\bm x}(\tau))d\tau,
          \qquad\qquad
       F^{[-2]}(t,{\bm x})\equiv
                    \int\limits_{-\infty}^{t}
                         F^{[-1]}(\tau,{\bm x}(\tau))d\tau\;.
\end{equation}
Integrals in equations (\ref{i1}) represent functions of time and
space coordinates. Integrals in equations (\ref{i2}) are defined
on the light-ray trajectory and are functions of time only.

Multiple time derivative from function $F(t,{\bm x})$ is denoted
by
\begin{equation}
\label{tder2} F^{(p)}(t,{\bm x})=\frac{\partial^p F(t,{\bm
x})}{\partial t^p}\;,
\end{equation}
so that its action on the time integrals eliminates integration in
the sense that
\begin{equation}
\label{tder3} F^{(p)}(t,{\bm
x})=\frac{\partial^{p+1}F^{(-1)}(t,{\bm x})}{\partial
t^{p+1}}=\frac{\partial^{p+2}F^{(-2)}(t,{\bm x})}{\partial
t^{p+2}}\;,
\end{equation}
and
\begin{equation}
\label{tder4} F^{[p]}(t,{\bm x})=\frac{d^{p+1}F^{[-1]}(t,{\bm
x})}{dt^{p+1}}=\frac{d^{p+2}F^{[-2]}(t,{\bm x})}{dt^{p+2}}\;.
\end{equation}

Spatial vectors will be denoted by bold italic letters, for
instance $A^i\equiv{\bm A}$, $k^i\equiv {\bm k}$, etc. The
Euclidean dot product between two spatial vectors ${\bm a}$ and
${\bm b}$ is denoted with dot between them: $a^ib_i={\bm
a}\cdot{\bm b}$. The Euclidean wedge (cross) product between two
vectors is denoted with symbol $\times$, that is
$\epsilon_{ijk}a^jb^k=({\bm a}\times{\bm b})^i$. Other particular
notations will be introduced as they appear in the text.

\section{Metric Tensor and Coordinate Systems}\label{gauge}
\subsection{Metric Tensor}
We assume that the gravitational field is weak everywhere along the light-ray
path. This allows us to work in the linear (with respect to $G$) approximation of General Relativity which is sometimes called the post-Minkowskian approximation \cite{damour}.
Thus, the metric tensor can be expressed as the sum of the Minkowski
metric and a small perturbation $h_{\alpha\beta}\ll1$ being proportional to the
universal gravitational constant $G$:
\begin{equation}
\label{m}
g_{\alpha\beta}=\eta_{\alpha\beta}+h_{\alpha\beta}.
\end{equation}
We consider propagation of light in the space-time around an
isolated gravitating system. The most general expressions for
metric perturbations in this case were given by Thorne
\cite{Tho80} in a special coordinate system which belongs to the
class of the harmonic coordinate systems. In what follows we shall
call these special coordinates \textit {canonical harmonic} coordinates. The
expressions given by Thorne read
\begin{align}
\label{bm} h_{00}^{\rm can.}=&\frac{2{\Mc}}{r}+
      2\sum_{l=2}^{\infty}\frac{(-1)^l}{l!}
       \left[\frac{{\Ic}_{A_l}(t-r)}{r}\right]_{,A_l} ,
\\
h_{0i}^{\rm can.}=& -\frac{2\epsilon_{ipq}{\Sc}_pN_q}{r^2}-
                 4\sum_{l=2}^{\infty}\frac{(-1)^ll}{(l+1)!}
          \left[\frac{\epsilon_{ipq}
                   {\Sc}_{pA_{l-1}}(t-r)}{r}\right]_{,qA_{l-1}}+\\
               &4\sum_{l=2}^{\infty}\frac{(-1)^l}{l!}
          \left[\frac{\dot{\mathcal
          I}_{iA_{l-1}}(t-r)}{r}\right]_{,A_{l-1}},\notag
\\
h_{ij}^{\rm can.}=&\delta_{ij}h_{00}^{\rm can.}+q_{ij}^{\rm can.},
\\
\label{em} q_{ij}^{\rm
can.}=&4\!\sum_{l=2}^{\infty}\!\!\frac{(-1)^l}{l!}\!\!
       \left[\frac{\ddot{\Ic}_{ijA_{l-2}}(t-r)}{r}
           \right]_{,A_{l-2}}\!\!\!\!-
        \!8\!\sum_{l=2}^{\infty}\!\frac{(-1)^ll}{(l+1)!}\!\!
         \left[\frac{\epsilon_{pq(i}\dot
          {\Sc}_{j)pA_{l-2}}(t-r)}{r}\right]_{,qA_{l-2}}.
\end{align}
Here $\Mc$ and $\Sc_i$ are the total mass and angular momentum of
the system, $\il $ and $\Sc_{A_l}$ are two independent sets of
mass-type and spin-type multipole moments, and $N^i=x^i/r$ is a
unit vector directed from the origin of the coordinate system to
the field point. Since the origin of the coordinate system has
been chosen at the center of mass, the expansions \eqref{bm} --
\eqref{em} do not depend on the mass-type linear multipole moment
$\Ic_i=0$.

The expressions for the metric perturbations in a different
coordinate system can be obtained by means of an infinitesimal
coordinate transformation
\begin{equation}
\label{trans} x'^\alpha=x^\alpha-w^\alpha
\end{equation}
from the \textit{canonical harmonic} coordinates $x^\alpha $ to
coordinates $x'^\alpha $ where $w^\alpha$ are gauge functions.
Then in the new coordinate system one has \cite{MTW73}
\begin{equation}
\label{met}
 h_{\alpha\beta}=h_{\alpha\beta}^{can.}+
         w_{\alpha,\beta}+
              w_{\beta,\alpha}\;.
\end{equation}
We note that if functions $w^\alpha$ satisfy the homogeneous wave
equation
\begin{equation}
\label{waveeq} \dAl w^\alpha=0,
\end{equation}
 the transformation \eqref{trans}
leaves us within the class of the harmonic coordinate systems, since
the harmonic gauge conditions can be formulated as $\dAl x^\alpha=0$ \cite{GrPo}.

The most general solution of the equation \eqref{waveeq} is given
in \cite{Tho80} (see also \cite{BD}) and reads
\begin{align}
\label{poh}
w^0=&\displaystyle{\sum_{l=0}^{\infty}} \left[\frac{{\cal
W}_{A_l}(t-r)}{r}\right]_{,A_l}\;,\\
\label{boh}
w^i=&\displaystyle{\sum_{l=0}^{\infty}} \left[\frac{{\cal
X}_{A_l}(t-r)}{r}\right]_{,iA_l}+\\
&\displaystyle{\sum_{l=1}^{\infty}}\biggl\{ \left[\frac{{\cal
Y}_{iA_{l-1}}(t-r)}{r}\right]_{,A_{l-1}}+
\left[\epsilon_{ipq}\frac{{\cal
Z}_{qA_{l-1}}(t-r)}{r}\right]_{,pA_{l-1}}\biggr\}\;,\notag
\end{align}
where ${\cal W}_{A_l}$, ${\cal X}_{A_l}$, ${\cal Y}_{iA_{l-1}}$,
and ${\cal Z}_{qA_{l-1}}$ are four sets of STF multipole moments
depending on the retarded time.

 We shall use the gauge freedom in choosing functions $w^\alpha$
to simplify the problems of integration of the equations of light
propagation and analysis of the observable effects. It turns out
that two coordinate systems possess properties useful for solution
of these problems. These are the harmonic and the
Arnowitt-Deser-Misner (ADM) coordinates.
\subsection{The Harmonic Coordinate System}
The harmonic coordinate system, analogous to the Lorentz gauge in
electrodynamics, is defined in linear approximation by the
conditions \cite{GrPo}
\begin{equation}
\label{gai} 2h^{\alpha\beta}_{\;\;\,,\beta}-h^{,\alpha}=0\;
\end{equation}
imposed on the metric perturbations, where $h\equiv h^\alpha_{\;\alpha}$. The linearized
Einstein field equations in harmonic coordinates reduce to wave
equations for metric perturbations $h^{\alpha\beta}$.

The metric tensor perturbations in harmonic coordinates \eqref{bm}
- \eqref{em} possess a property that can be used to simplify the
problem of analytical integration of the equations of light
propagation. Specifically, the method of integration of equations
of light geodesics, developed first in \cite{Kop97} for stationary fields, and extended in \cite{KSGU99} to time-dependent fields,
is utilized in the present work (Section \ref{methsol}). This method takes advantage of the fact
that the metric perturbations are functions of the retarded time.
Not all coordinates in general relativity have this property. For instance, metric
perturbations in the canonical ADM coordinate system, considered
in the next section, contain terms that depend on instantaneous
time, which does not allow us to use the abovementioned method of integration with the retarded time.
\subsection{The Arnowitt-Deser-Misner Coordinate System}
The ADM gauge conditions \cite{ADM62} in the linear approximation
read
\begin{eqnarray}
\label{admgauge}
2h_{0i,i} - h_{ii,0} = 0\;, \quad \quad 3h_{ij,j} - h_{jj,i} =
0\;.
\end{eqnarray}
For comparison, the harmonic gauge conditions \eqref{gai} in the
linear approximation have the form
\begin{eqnarray}
\label{harmonic}
2h_{0i,i} - h_{ii,0} = h_{00,0}\;, \quad \quad 2h_{ij,j} -
h_{jj,i} = - h_{00,i}\;.
\end{eqnarray}

The ADM coordinates have a property that test particles perturbed
by gravitational waves stay at rest with respect to these
coordinates, that is the ADM coordinate system is co-moving with test particles. This
property can be used to simplify analysis of observable effects.
For example, when one considers detection of a photon by an
observer in the field of gravitational wave, then normally one
would have to solve both the equations of motion of the photon and
observer \cite{GrPo}. Using ADM coordinates allows us to exclude the problem of
motion of the observer from the coordinate description of the gravitational-wave effect on the photon.

The disadvantage of using the ADM coordinates for analysis of the
effects in propagation of light through gravitational field of an
isolated source is that, as it was mentioned above, the ADM metric
perturbations normally contain terms that are functions of
instantaneous time. This does not allow one to use mathematical advantages of the method of
integration of the equations of light propagation (Section
\ref{methsol}).

Fortunately, the classes of ADM and harmonic coordinates overlap.
In \cite{KSGU99} an \textit{ADM-harmonic} coordinate system was
constructed in the case when isolated system possesses constant
mass and angular momentum and time-dependent quadrupole moment. We
generalize this result to the case of isolated system possessing
the whole multipole structure \eqref{bm}-\eqref{em}. The gauge
functions
\begin{align}
\label{w0}
       w^0=&\sum_{l=2}^{\infty}\ffrac{(-1)^l}{l!}
            \left[\frac{
            {\Ic}^{(-1)}_{A_l}(t-r)}{r}\right]_{,A_l},
\\
               w^i=& \sum_{l=2}^{\infty}\ffrac{(-1)^l}{l!}
\label{wi}
            \left[\frac{
            {\Ic}^{(-2)}_{A_l}(t-r)}{r}\right]_{,iA_l}-
4\sum_{l=2}^{\infty}\ffrac{(-1)^l}{l!}
                 \left[\ffrac{{\Ic}_{iA_{l-1}}(t-r)}{r}
                                    \right]_{,A_{l-1}}+
\\
&4\sum_{l=2}^{\infty}\ffrac{(-1)^ll}{(l+1)!}
              \left[\ffrac{\epsilon_{iba}
                             {\mathcal
                             S}^{(-1)}_{bA_{l-1}}(t-r)}{r}\right]_{,aA_{l-1}}\notag
\end{align}
bring the metric perturbations $h_{\alpha\beta}$ to the form
\begin{eqnarray}\label{adm1}
h_{00}&=&\frac{2{\cal M}}{r}\;,\\
\nonumber\\\label{adm2}
h_{0i}&=&-\frac{2\epsilon_{ipq}{\cal S}_p N_q}{r^2}\;,\\
\nonumber\\\label{adm3}
h_{ij}&=&\delta_{ij}h_{00}+h^{TT}_{ij}\;,\\
\nonumber\\\label{adm4} h^{TT}_{ij}&=& P_{ijkl}q_{kl}^{\rm
can.}\;,
\end{eqnarray}
which satisfies both harmonic and ADM gauge conditions. Here the
TT-projection differential operator $P_{ijkl}$, applied to
symmetric tensors depending on both time and spatial coordinates,
is given by
\begin{eqnarray}
P_{ijkl}&=&(\delta_{ik} - \Delta^{-1} \partial_i  \partial_k)
              (\delta_{jl} - \Delta^{-1} \partial_j  \partial_l) -
\frac{1}{2} (\delta_{ij} - \Delta^{-1} \partial_i  \partial_j)
              (\delta_{kl} - \Delta^{-1} \partial_k  \partial_l)\;,
\end{eqnarray}
and $\Delta$ and $\Delta^{-1}$  denote the Laplacian and the
inverse Laplacian respectively.

The \textit{ADM-harmonic} coordinates combine the advantages of
both coordinate systems and will be used in what follows for
analysis of observable effects in propagation of electromagnetic
signals through the gravitational field of the isolated system.
\section{Equations for Propagation of Electromagnetic Signals}
\subsection{Geometrical Optics in Curved Space-Time}
In this section following \cite{KoMa02} and \cite{MTW73} (see \S 22.5)
we provide a derivation of the main laws of geometrical optics in
curved space-time from the Maxwell's equations.

In absence of sources the Maxwell's equations for the
electromagnetic field tensor $F_{\alpha\beta}$ in curved
space-time take the well known form \cite{LaLi62, MTW73}
\begin{equation}
\label{m1a}
\nabla_\alpha F_{\beta\gamma}+\nabla_\beta
F_{\gamma\alpha}+\nabla_\gamma F_{\alpha\beta}=0\;,
\end{equation}
\begin{equation}
\label{m1b}
\nabla_\beta F^{\alpha\beta}=0\;,
\end{equation}
where $\nabla_\alpha$ denotes a covariant derivative. The wave
equation for $F_{\alpha \beta}$ can be derived from \eqref{m1a}
and \eqref{m1b} and in vacuum ($R_{\alpha\beta}=0$, where
$R_{\alpha\beta}$ is the Ricci tensor) it assumes the form
\begin{equation}
\label{m2} {\dAl}_g
F_{\alpha\beta}+R_{\alpha\beta\mu\nu}F^{\mu\nu}=0\;,
\end{equation}
where $\dAl_g\equiv \nabla^\alpha\nabla_\alpha$ and
$R_{\alpha\beta\gamma\delta}$ is the Riemann curvature tensor of
the space-time.

The solution to the Maxwell's equations corresponding to a high
frequency wave is given by
\cite{BKNP1,BKNP2}
\begin{equation}
\label{ff1} F_{\alpha\beta}={\rm
Re}\{A_{\alpha\beta}\exp(i\varphi)\}\;,
\end{equation}
where $A_{\alpha\beta}$ is the complex amplitude, which is a
slowly varying function of position and time, and $\varphi$ is the
phase, rapidly changing with position and time.

The criterion for applicability of the geometrical optics
approximation in curved space-time is formulated \cite{MTW73}
as follows \cite{MTW73}. Let $\mathcal R$ be the characteristic radius of
curvature of the space-time through which the electromagnetic wave
propagates and $\mathcal L$ -- the characteristic length, over
which the amplitude, wavelength and polarization of the
electromagnetic wave change significantly. Geometrical optics
approximation can be applied whenever the wavelength of the
electromagnetic radiation satisfies the following two conditions:
1)$\lambda\ll \mathcal R$; and 2)$\lambda \ll \mathcal L$. Then a
small parameter $\varepsilon\equiv\lambda/min\{\mathcal L,
\mathcal R\}$ can be introduced, and the expansion of the
electromagnetic field of a wave \eqref{ff1} in powers of
$\varepsilon$ is assumed to have the form
\begin{equation}
\label{exp} F_{\alpha\beta}=\left(a_{\alpha\beta}+ \varepsilon
b_{\alpha\beta}+\varepsilon^2
c_{\alpha\beta}+...\right)\exp\left(\frac{i\varphi}{\varepsilon}\right)\;.
\end{equation}
For further discussion of this procedure see \cite{Mash87} and
\cite {PeHa93}. Substituting this expansion into \eqref{m1a},
taking into account the definition of the electromagnetic wave
vector $l_\alpha\equiv\partial \varphi/\partial x^\alpha$ and
rearranging the terms with the same powers of $\varepsilon$ lead
to the chain of equations
\begin{align}
\label{m4a} l_\alpha\; a_{\beta\gamma}+l_\beta\;
a_{\gamma\alpha}+l_\gamma\;
a_{\alpha\beta}=&\;0\;,\\
\label{m4b} \nabla_\alpha\; a_{\beta\gamma}+\nabla_\beta\;
a_{\gamma\alpha}+\nabla_\gamma\; a_{\alpha\beta}=&-i\left(
l_\alpha\; b_{\beta\gamma}+l_\beta\; b_{\gamma\alpha}+l_\gamma\;
b_{\alpha\beta}\right)\;,
\end{align}
where the effects of curvature appears in the next approximation and, hence, have been neglected. Another chain
of equations is obtained by substituting \eqref{exp} into
\eqref{m1b}:
\begin{eqnarray}
\label{m5a}
l_\beta\; a^{\alpha\beta}&=&0\;,\\
\label{m5b} \nabla_\beta\; a^{\alpha\beta}+il_\beta\;
b^{\alpha\beta}&=&0\;.
\end{eqnarray}
The equation \eqref{m5a} shows that in the lowest order
approximation (terms of the order $1/\varepsilon$) the amplitude
of the electromagnetic field tensor is perpendicular to the wave
vector in the four dimensional sense. Contracting \eqref{m4a} with
$l_\alpha$ and accounting for \eqref{m5a} implies that $l_\alpha$
is a null vector
 \begin{equation}
\label{null}
 l_\alpha l^\alpha=0.
 \end{equation}
 Taking a covariant derivative from the equality \eqref{null} and using the fact that
$\nabla_{[\alpha}l_{\beta]}=0$ (since
$l_\alpha\equiv\nabla_\alpha\varphi$) we obtain that $l_\alpha$
satisfies the geodesic equation
\begin{equation}
\label{p0} {l}^\beta\nabla_\beta{l}^\alpha=0\;,
\end{equation}
showing that the wave vector is parallel transported along itself.
Equations \eqref{null} and \eqref{p0} combined constitute an
important result of the geometrical optics: in the lowest order
approximation light rays are null geodesics.

Substituting \eqref{ff1} into \eqref{m2} and considering terms of
the order $1/\varepsilon$ one can obtain the law of propagation
for the amplitude of the electromagnetic tensor
\begin{equation}
\label{m6} D_\lambda a_{\alpha\beta}+\vartheta
a_{\alpha\beta}=0\;,
\end{equation}
where $D_\lambda\equiv l^\alpha\nabla_\alpha$ and $\vartheta\equiv
(1/2) \nabla_\alpha l^\alpha$ is the expansion of the null
congruence $l^\alpha$.
\subsection{Equations for Trajectory of a Photon}
It was shown in the previous section that in the lowest order
approximation of geometrical optics light rays, or trajectories of
photons, are represented by null geodesics of the space-time under
consideration. In what follows we shall consider propagation of
photons in a space-time around an isolated gravitating system.
With the metric of the space-time defined by \eqref{met}. We shall
assume that the wavelength of the electromagnetic radiation is
much smaller than the characteristic wavelength of the
gravitational waves emitted by the gravitating system, so that the
conditions for validity of geometrical optics approximation are
satisfied.

Let us consider propagation of photons subject to the
initial-boundary conditions
\begin{equation}\label{ibc}
      {\bm x}(t_0)= {\bm x}_0 ,\qquad
     \frac {d {\bm x}(-\infty)}{dt}= {\bm k}\;,
 \end{equation}
which specify the spacial velocity of photons at the past null
infinity by a unit spacial vector $\bf k$  and the positions of
the photons at some initial instant of time $t_0$.

Taking into account that $l^\alpha=dx^\alpha/d\lambda$, where
$\lambda$ is an affine parameter along the photon's trajectory, we
can rewrite the equations of geodesics \eqref{p0} in terms of
coordinates of a photon
\begin{equation}
\label{gpa}
\frac{d^2{x}^\alpha}{d\lambda^2}+\Gamma^\alpha_{\beta\gamma}{\frac{d{x}^\beta}{d\lambda}}{\frac{dx^\gamma}{d\lambda}}=0\;,
\end{equation}
where
\begin{equation}
\label{bb1}
\Gamma^\alpha_{\beta\gamma}=\frac{1}{2}g^{\alpha\mu}\left(
\partial_\gamma g_{\mu\beta}+\partial_\beta g_{\mu\gamma}-
\partial_\mu g_{\beta\gamma}\right)
\end{equation}
are the Christoffel symbols. Combining the four equations
\eqref{gpa} we obtain the equations for spacial position of a
photon as a function of the coordinate time
\begin{equation}
\label{gpt1} \ddot
{x}^i(t)=-\left(\Gamma^i_{\alpha\beta}-\Gamma^0_{\alpha\beta}\dot
x^i\right)\dot x^\alpha\dot x^\beta\;.
\end{equation}
Substituting \eqref{bb1} into \eqref{gpt1}, taking into account
that $\dot x^i=k^i+O(h)$ and keeping only linear in metric
perturbations terms we can rewrite the equations \eqref{gpt1} in
the form
\begin{align}
\label{eq1}
 \ddot{x}^i(t)=&\frac{1}2 h_{00,i}-
                          h_{0i,0}-
                  \frac{1}2 h_{00,0}k^i-
                             h_{ik,0}k^k-
                            ( h_{0i,k}-
                               h_{0k,i})k^k-\\
\nonumber
              &  h_{00,k} k^kk^i-
            \left( h_{ik,j}-
           \frac{1}2 h_{kj,i}\right)k^kk^j+
       \left(\frac{1}2h_{kj,0}-
                        h_{0k,j}\right)k^kk^jk^i\;.
\end{align}
In section \ref {sol} we will show how these equations can be
analytically integrated in the space-time with metric \eqref{met}
defining gravitational field of an isolated gravitating system.
\subsection{Equations for Propagation of Polarization Properties}
In this section, following \cite{AnBr74} and \cite{KoMa02}, we
introduce the relativistic description of polarized
electromagnetic radiation and give the equation (derived
in~\cite{KoMa02}) for gravity-induced rotation of the polarization
plane of an electromagnetic wave (the Skrotskii effect).

For description of the polarization properties of electromagnetic
radiation it is necessary to introduce local reference frames
along the light rays. At each point of space-time we introduce a
complex null tetrad $(l^\alpha,n^\alpha,m^\alpha,\overline
m^\alpha)$ associated with a congruence of light rays
\cite{NePe62, Frol77}. All the vectors of the tetrad are
lightlike. The vectors $l^\alpha$ and $n^\alpha$ are real; vectors
$m^\alpha$ and $\overline m^\alpha$ are complex and complex
conjugate with respect to each other. The tetrad is normalized in
such a way that the only nonvanishing products among the vectors
are $l_\alpha n^\alpha=-1$ and $m_\alpha\overline m^\alpha=1$.

It is obvious that such field of complex null tetrads is not
uniquely determined by the congruence of null rays and the
normalization conditions. The transformations
\begin{align}
\label{tetr} l'^\alpha&=Al^\alpha\;,
\\
 n'^\alpha&=A^{-1}\left(n^\alpha+\overline{B}
m^\alpha+B\overline{m}^\alpha+B\overline{B} l^\alpha\right)\;,
\nonumber\\
m'^\alpha&=e^{-i\Theta}\left(m^\alpha+\overline{B}l^\alpha\right)\;,
\nonumber\\
\overline{m}'^\alpha&=e^{i\Theta}\left(\overline{m}^\alpha+Bl^\alpha\right)\;,
\nonumber
\end{align}
where $A$ and $\Theta$ are real and $B$ is a complex parameter,
preserve the direction of the vector $l^\alpha$ and the
normalization conditions \cite{Frol77}. The transformations
\eqref{tetr} form a four-parameter subgroup of the Lorentz group.
 In addition to the complex null tetrad we introduce at each
point in space-time an orthonormal reference tetrad
$e^\alpha_{\;(\beta)}$ defined as follows. Suppose at each point
of space-time there is an observer moving with four-velocity
$u^\alpha$. Let two of the vectors of the observer's local
reference frame be
\begin{equation}
\label{p3} e^\alpha_{\;(0)}=u^\alpha\;,\qquad\quad
e^\alpha_{\;(3)}=(-{l}_\alpha u^\alpha)^{-1}\left[{l}^\alpha+
({l}_\beta u^\beta)u^\alpha\right]\;.
\end{equation}
With such orientation of the reference frame the observer will see
the electromagnetic wave propagating in the $+z$ direction.
Spacelike unit vectors $e^\alpha_{\;(1)}$ and $e^\alpha_{\;(2)}$
are orthogonal to each other as well as to both $e^\alpha_{\;(0)}$
and $e^\alpha_{\;(3)}$ and thus are specified up to a spacial
rotation.

The relationship between the vectors of the complex null tetrad
and the frame $e^\alpha_{\;(\beta)}$ is given by
\begin{align}
\label{etn}
l^\alpha&=-(l_\gamma u^\gamma)\left(e^\alpha_{\;(0)}+e^\alpha_{\;(3)}\right)\;,\\
n^\alpha&=-\frac{1}{2}(l_\gamma u^\gamma)\left(e^\alpha_{\;(0)}-e^\alpha_{\;(3)}\right)\;,\\
m^\alpha&=\frac{1}{\sqrt{2}}\left(e^\alpha_{\;(1)}+ie^\alpha_{\;(2)}\right)\;,\\
\overline{m}^\alpha&=\frac{1}{\sqrt{2}}\left(e^\alpha_{\;(1)}-ie^\alpha_{\;(2)}\right)\;.
\end{align}
Vectors $e^\alpha\;_{(1)}$ and $e^\alpha\;_{(2)}$ and
corresponding to them $m^\alpha$ and $\overline m^\alpha$ play an
important role in description of polarization properties of
electromagnetic radiation since they form a basis in the
polarization plane.

The tensor of the electromagnetic field can be represented as
\begin{equation}
\label{fab} F_{\alpha\beta}=Re({\cal F}_{\alpha\beta})\;,
\end{equation}
where ${\cal F}_{\alpha\beta}$ is the complex field. The complex
field can be expressed as
\begin{equation}
\label{fabc}
{\cal F}_{\alpha\beta}=\Phi
l_{[\alpha}m_{\beta]}+\Psi l_{[\beta}\overline m_{\alpha]}
\end{equation}
 where $\Phi$ and $\Psi$ are complex scalar functions.

The electric and magnetic fields in the rest frame of an observer
moving with four-velocity $u^\alpha$  are defined as
\begin{align}
\label{emf} E^\alpha=F^{\alpha\beta}u_\beta\;,\quad
H^\alpha=(-1/2)\epsilon^{\alpha\beta\gamma\delta}F_{\gamma\delta}u_\beta\;.
\end{align}
We also define the complex electric field as
\begin{equation}
\label{elfc}
{\cal E}^\alpha={\cal F}^{\alpha\beta}u_\beta\;.
\end{equation}
Polarization properties of light are completely characterized by
the polarization tensor~\cite{LaLi62} (coherency matrix in
\cite{BoWo99})
\begin{equation}
\label{jab}
J_{\alpha\beta}=\langle {\cal E}_\alpha {\cal \overline
E}_\beta\rangle\;,
\end{equation}
where the angular brackets notify an ensemble average equivalent
to averaging over many periods of the wave.

The electromagnetic Stokes parameters are defined with respect to
two of  the observer's tetrad vectors $e^\alpha_{\;(1)}$ and
$e^\alpha_{\;(2)}$ as follows \cite{AnBr74}
\begin{align}
\label{stp0}
S_0&=J_{\alpha\beta}\left[e^\alpha_{\;(1)}e^\beta_{\;(1)}+
e^\alpha_{\;(2)}e^\beta_{\;(2)}\right]\;,\\
\label{stp1}
S_1&=J_{\alpha\beta}\left[e^\alpha_{\;(1)}e^\beta_{\;(1)}-
e^\alpha_{\;(2)}e^\beta_{\;(2)}\right]\;,\\
\label{stp2}
S_2&=J_{\alpha\beta}\left[e^\alpha_{\;(1)}e^\beta_{\;(2)}+
e^\alpha_{\;(2)}e^\beta_{\;(1)}\right]\;,\\
\label{stp3}
S_3&=iJ_{\alpha\beta}\left[e^\alpha_{\;(1)}e^\beta_{\;(2)}-
e^\alpha_{\;(2)}e^\beta_{\;(1)}\right]\;.
\end{align}
It is worth noting that both the Stokes parameters and the
components of the polarization tensor of an electromagnetic wave
can be determined from simple experiments by measuring the
intensities of the wave after it passes through devices that
transmit only radiation of certain polarization \cite{BoWo99}.

Using the definition of the polarization tensor \eqref{jab} we can
also rewrite the expressions for the Stokes parameters in terms of
the electric field components
\begin{align}
\label{stpf0}
S_0&=<|{\cal E}_{(1)}|^2+|{\cal E}_{(2)}|^2>\;,\\
\label{stpf1}
S_1&=<|{\cal E}_{(1)}|^2-|{\cal E}_{(2)}|^2>\;,\\
\label{stpf2}
S_2&=<{\cal E}_{(1)}\overline{\cal E}_{(2)}+\overline{\cal E}_{(1)}{\cal E}_{(2)}>\;,\\
\label{stpf3}
S_3&=i<{\cal E}_{(1)}\overline{\cal E}_{(2)}-\overline{\cal E}_{(1)}{\cal E}_{(2)}>\;,
\end{align}
where ${\cal E}_{(n)}={\cal E}_\alpha e^\alpha_{\;(n)}$ for
$n=1,2$. The Stokes parameters can also be expressed in terms of
the complex functions $\Phi$ and $\Psi$ since
\begin{equation}
\label{ftoe} {\cal E}_\alpha=-\frac{l^\alpha
u_\alpha}2\left(\Phi\; m_\alpha+\Psi\; \overline m_\alpha\right)
\end{equation}
and therefore
\begin{equation}
\label{ftoe12} {\cal E}_{(1)}=-\frac{l^\alpha u_\alpha}{\sqrt
8}\left(\Phi+\Psi\right)\;,\quad {\cal E}_{(2)}=-i\frac{l^\alpha
u_\alpha}{\sqrt 8}\left(\Phi+\Psi\right)\;.
\end{equation}

 To determine how the polarization
properties of light vary along the ray we should specify the law
of propagation for the tetrad vectors.
 The vectors of both the null tetrad and the tetrad
$e^\alpha_{\;(\beta)}$ are postulated to be parallelly transported
along the light rays. We specify the tetrads at the point of
observation and thus specify them at every point along the ray.
Thus, the equations for propagation of the tetrad vectors read
\begin{align}
\label{prm}
\frac{dm^\alpha}{d\lambda}+\Gamma^\alpha_{\beta\gamma}\,l^\beta\,
m^\gamma&=0\;,\\
\label{prmc}
\frac{d\overline{m}^\alpha}{d\lambda}+\Gamma^\alpha_{\beta\gamma}\,l^\beta\,
\overline{m}^\gamma&=0\;,\\
\label{pre}
\frac{de^\alpha_{\;(\mu)}}{d\lambda}+\Gamma^\alpha_{\beta\gamma}\,l^\beta\,
e^\gamma_{\;(\mu)}&=0\;,
\end{align}
and the same laws hold for vectors $l^\alpha$ and $n^\alpha$ of
the null tetrad.

It follows from \eqref{m6}, \eqref{elfc} and \eqref{pre} that the
complex amplitude of the electric field propagates along the ray
according to the law
\begin{equation}
\label{prcf} D_\lambda {\cal E}_\alpha+\vartheta{\cal
E}_\alpha=0\;.
\end{equation}
For the complex amplitudes $\Phi$ and $\Psi$ one has
\begin{equation}
\label{prpp} D_\lambda \Phi+\vartheta\Phi=0\;,\quad D_\lambda
\Psi+\vartheta\Psi=0\;.
\end{equation}
Finally, from the equations \eqref{stpf0} - \eqref{stpf3} it
follows that the Stokes parameters propagate according to the law
\begin{equation}
\label{prsp} D_\lambda S_\alpha+2\vartheta S_\alpha=0\;,
\end{equation}
where $\alpha = 0,1,2,3.$ We would like to note that despite
suggestive notation the Stokes parameters do not form a
four-vector, because they do not behave as a vector under
coordinate system transformations.

Any stationary or time-dependent axisymmetric gravitational field
in general causes a relativistic effect of rotation of the
polarization plane of an electromagnetic wave \cite{KoMa02}. This
effect was first discussed by Skrotskii \cite{Skr57} and
afterwards by many researchers (see, for example, \cite {Mash75}
and references therein). Recently the effect was studied in
\cite{KoMa02} where authors derived an expression for the angle of
rotation of the polarization plane of an electromagnetic  wave
propagating through a weak gravitational field described by metric
perturbations $h_{\alpha\beta}$. Here we give this expression
without derivation which can be found in \cite{KoMa02} or in
\cite{KoKo05}. If $\phi$ is the characterizing the orientation of
the polarization ellipse, then one has
\begin{equation}
\label{skr}
 \frac{d\phi}{d t}=\frac{1}2
        k^{\alpha}k^j
            \epsilon_{j\hat{p}\hat q}\partial_q
                h_{\alpha\hat{p}}\;,
\end{equation}
where the hat over the spatial indices denotes the projection onto
the plane orthogonal to the propagation of light ray, for
instance, $A^{\hat{i}}\equiv P^i_{\;j}A^j$. In the following
section we shall integrate this equation analytically to obtain
the angle of the rotation of the polarization plane for an
electromagnetic signal propagating in the gravitational field of
an isolated system.
\section{Solution to the Equations of Propagation of Electromagnetic Signals}\label{sol}
\subsection{Method of Analytical Integration of the Equations of
Propagation of Electromagnetic Signals}\label{methsol} In this
section we describe the method of analytical integration of
equations of light propagation in the field of an isolated
gravitating system developed in the series of publications \cite
{Kop97}, \cite{KSGU99} and \cite{KoKo05}. In subsequent sections
this method will be applied to integrate the equations of
geodesics \eqref{eq1} and the equation for Skrotskii effect
\eqref{skr} in the space-time with the metric given by \eqref{met}
around an isolated gravitating system.

We introduce new variables $\tau$ and $\xi^i$ as follows
\begin{equation}
\label{nvar} \
\tau\equiv k_ix^i\;,\qquad \xi^i\equiv P^i_{\;j}x^j\;,
\end{equation}
where $P^i_{\;j}$ is the operator of projection onto the plane
perpendicular to $k^i$. If one considers unperturbed trajectory of
a photon $x^i_N=k^i(t-t_0)+x_0$, it is easy to see that the
variable $\tau$ characterizing the position of the photon is
proportional to time
\begin{equation}\label{tt*}\tau\equiv
k_ix^i_N=t-t^*,
\end{equation}
 where $t^*\equiv k_i x^i_0-t_0$
is the time of the closest approach of the electromagnetic signal to the origin of the coordinate system.
Since $t^*$ is a constant for a particular ray, one has $d\tau=dt$. This allows to change to
the variable $\tau$ when calculating  integrals along the unperturbed ray.  Vector $\xi^i$ is the vector
from the origin of the coordinate system to the point of the closest approach. In terms of the new variables the
unperturbed trajectory can be written as
\begin{equation}
\label{unt2}
x^i_N(\tau)=k^i\tau+\xi^i\;.
\end{equation}
Since the vectors $k^i$ and $\xi^i$ are orthogonal to each other,
the distance from the point on unperturbed trajectory with
coordinates $\tau$ and $\xi^i$ to the origin of the coordinate
system can be expressed as
\begin{equation}
\label{rtd}
r_N=\sqrt{\tau^2+d^2}\;,
\end{equation}
where $d=|{\bm \xi}|$ is the impact parameter of the unperturbed
light-ray trajectory with respect to the origin of the coordinate
system.

 We introduce the operators of
differentiation with respect to $\tau$ and $\xi^i$
\begin{equation}
\label{do}
\hat\partial_\tau\equiv\frac{\partial}{\partial\tau}\;,\qquad
\hat\partial_i\equiv P_i^{\;j}\frac{\partial}{\partial\xi^j}\;.
\end{equation}
Then for any smooth function $F(t,x^i)$ the following relationships are valid between
the derivatives in the old and new variables taken on the unperturbed trajectory
\begin{align}
\label{rdif1}
\left[\left(\frac{\partial}{\partial x^i}+
       k_i\frac{\partial}{\partial t}\right)F(t\,,\,{\bm x})
        \right]_{{\bm x}= {\bm x}_0+{\bm k}(t-t_0)}      =&
    \left(\frac{\partial}{\partial\xi^i}+
       k_i\frac{\partial}{\partial\tau}\right)
        F(t^*+\tau\,,\,{\bm \xi}+{\bm k}\tau)\;,
\\
 \label{rdif2}
\left[\frac{\partial}{\partial t}F(t\,,\,{\bm x})\right]_{{\bm
x}={\bm x}_0+ {\bm k}(t-t_0)}=&\frac{\partial}{\partial t^*}
F(t^*+\tau\,,\,{\bm \xi}+{\bm k}\tau).
\end{align}
In the left-hand sides of the equations \eqref{rdif1} and \eqref{rdif2} one has to first calculate the
derivatives and only after that substitute the unperturbed trajectory ${\bm x}= {\bm x}_0+{\bm k}(t-t_0)$ while
in the right hand sides one has to first substitute the unperturbed trajectory parameterized by the variables $\tau$ and
$\xi^i$ and then differentiate. Using the equation \eqref{rdif2} in \eqref{rdif1} one can obtain the expression
for spacial derivatives
\begin{equation}
\label{rdif3}
\left[\frac{\partial F(t,{\bm x})}{\partial x^i}\right]_{{\bm x}= {\bm x}_0+{\bm k}(t-t_0)}
       =
    \left(\frac{\partial}{\partial\xi^i}+
       k_i\frac{\partial}{\partial\tau}-k_i\frac{\partial}{\partial t^*}\right)
        F(t^*+\tau\,,\,{\bm \xi}+{\bm k}\tau)\;.
\end{equation}
Equations \eqref{rdif2} and \eqref{rdif3} can be used for changing over to the variables $\tau$ and $\xi^i$ in the equations
of light propagation.
A useful property of the variables $\tau$  and $\xi^i$ is that
when one calculates integrals along the light rays the following
formulae are valid
\begin{align}
\label{ip1}
             \int\frac{\partial}{\partial\tau}F(\tau,{\bm \xi})\,d\tau
                                             =&F(\tau,{\bm \xi})
                                +C({\bm\xi})\;,\\\nonumber\\
\label{ip2} \int\frac{\partial}{\partial\xi^i}F(\tau,{\bm
\xi})\,d\tau
  =&\frac{\partial}{\partial\xi^i}\int F(\tau,{\bm \xi})\,d\tau\;,
\end{align}
where $C({\bm\xi})$ is a function of ${\bm \xi}$. Equation
\eqref{ip1} shows that the terms represented as partial
derivatives with respect to $\tau$ can be immediately integrated.
Equation \eqref{ip2} states that one can change the order of
integration and differentiation which crucially simplifies the
problem of integration of the equations as we show below.

Let us consider the geodesic equations \eqref{eq1}. All terms in
the right-hand sides of the equations  are proportional to the
first-order derivatives of the metric perturbations
$h_{\alpha\beta}$ with respect to time and spacial coordinates.
The metric perturbations, given by the equations \eqref{met},
consist of the canonical and gauge-dependent parts. The canonical
part (Eqns. \eqref{bm}-\eqref{em}) is expressed as a linear
combination of derivatives of different orders with respect to
spacial variables of functions $\left[F(t-r)/r\right]$ (where
$F(t-r)$ denotes a mass- or spin-type multipole moment of the
system). If one changes to the variables $\tau$ and $\xi^i$ in the
geodesic equations \eqref{eq1}, integrates and changes the order
of differentiation and integration in the right-hand sides of the
equations, then the only two types of integrals that will appear
in the solutions (for a moment, leave apart the terms produced by
the gauge-dependent components in the metric perturbations) will be the
integrals of the type $[F(t-r)/r]^{[-1]}$ and $[F(t-r)/r]^{[-2]}$.
The gauge-dependent terms, after changing the variables to $\tau$
and $\xi^i$, will appear in the equations of geodesics under the
second-order derivative with respect to $\tau$ and thus can be
immediately integrated (cf. Eq. \eqref{ip2}). A similar
consideration of the integration procedure can be done for the equations describing
the Skrotskii effect.

As it follows from the consideration above, the problem of
integration of the equations of light propagation reduces to
evaluation of integrals $[F(t-r)/r]^{[-1]}$ and
$[F(t-r)/r]^{[-2]}$ where $F(t-r)$ denotes  multipole moments of
different type and order of the gravitating system.

Let us  consider first the contributions to the relativistic
effects in propagation of light due to the mass-monopole and
spin-dipole terms. We neglect the loss of energy and angular
momentum due to gravitational radiation so that the total mass and
angular momentum of the system are considered to be conserved.
Then it can be shown that the contributions under consideration
are expressed in terms of integrals $\left[1/r\right]^{[-1]}$,
$\left[\dksi_i(1/r)\right]^{[-1]}$,
$\left[\dksi_{ia}(1/r)\right]^{[-1]}$,
$\left[\dksi_i(1/r)\right]^{[-2]}$ and
$\left[\dksi_{ia}(1/r)\right]^{[-2]}$ which can be evaluated.

The relativistic corrections to the light ray path and the
relativistic effect of rotation of the polarization plane due to
higher order multipole moments (starting with the mass- and
spin-quadrupole) are expressed in terms of integrals
$[F(t-r)/r]^{[-1]}$ and $[F(t-r)/r]^{[-2]}$ along the unperturbed
light ray, where $F(t-r)$ denotes a multipole moment. To evaluate
these integrals we introduce the new variable~\cite{KSGU99}
\begin{equation}
\label{y}
  y\equiv s-t^*=\tau-r(\tau)=\tau-\sqrt{d^2+\tau^2}\;.
\end{equation}
Then the following relationships hold
\begin{equation}
\label{dy} \tau=\frac{y^2-d^2}{2y}\;, \qquad
\sqrt{d^2+\tau^2}=-\frac{1}2\frac{d^2+y^2}y\;, \qquad
d\tau=\frac{1}2\frac{d^2+y^2}{y^2}dy.
\end{equation}
Making use of the new variable $y$ and the equations \eqref{dy} we
can express the integrals under discussion as follows
\begin{align}
\label{ins1} \left[\frac{F(t-r)}r\right]^{[-1]}=& -
\int\limits_{-\infty}^{y}
          \frac{F(t^*+\zeta)}\zeta
  d\zeta\;,
\\
\label{ins2} \left[\frac{F(t-r)}r\right]^{[-2]}=& -\frac{1}2
  \int\limits_{-\infty}^{y}
   \int\limits_{-\infty}^{\eta}\frac{F(t^*+\zeta)}\zeta
    d\zeta
     d\eta
-\frac{d^2}2
      \int\limits_{-\infty}^{y}
    \!\!\!\frac{1}{\eta^2}\!\!\!
        \int\limits_{-\infty}^{\eta}\frac{F(t^*+\zeta)}\zeta
    d\zeta
     d\eta\;,
\end{align}
where $\zeta$, $\eta$ are new variables of integration, and $t^*$
is the time of the closest approach of photon to the origin of the
coordinate system. It is worth to note that the time $t^*$ depends
on the choice of coordinate system and therefore in general has no
physical meaning.

An important property of the integrals \eqref{ins1} and
\eqref{ins2}, expressed in terms of the variable $y$, is that they
depend on $\tau$ and $\xi^i$ only through the upper limits of
integration (since $y\equiv\tau-\sqrt{d^2+\tau^2}$) and the square
of the impact parameter $d^2$ in the prefactor of the second
integral in \eqref{ins2}.

Differentiating \eqref{ins1} with respect to $\xi^i$ and $\tau$
yields
\begin{align}
\label{d11}
\dksi_i\left\{\left[\frac{F(t-r)}r\right]^{[-1]}\right\}&=
-\frac{F(t^*+y)}y\;\dksi_iy=-F(t^*+y)\;\dksi_i\ln(-y)= \frac{\xi^i}{yr}\,F(t-r)\;,\\
\label{dt1}
\hat\partial_\tau\left\{\left[\frac{F(t-r)}r\right]^{[-1]}\right\}&=-\frac{F(t^*+y)}y\;\hat\partial_\tau
y=-\frac{F(t^*+y)}y\left(1-{\frac{\tau}r}\right)=\frac{F(t-r)}r\;.
\end{align}
The last result also follows from the equation \eqref{ip2}. Thus
derivatives of the integral $[F(t-r)/r]^{[-1]}$ with respect to
either $\xi^i$ or $\tau$ can be expressed in terms of the
integrand.

Differentiating the \eqref{ins2} with respect to $\xi^i$ multiple
times yields
\begin{align}
\label{d12}
  \dksi_k\left\{
   \left[\frac{F(t-r)}r\right]^{[-2]}\right\}=&
\xi^k
      \left\{\frac{1}y\left[\frac{F(t-r)}r\right]^{[-1]}-
      \int\limits_{-\infty}^{y}
     \frac{1}{\eta^2}
         \int\limits_{-\infty}^{\eta}\frac{F(t^*+\zeta)}\zeta\,
    d\zeta
     d\eta
     \right\}\;,
\\
\label{d22}
 \dksi_{jk}\left\{\left[\frac{F(t-r)}r\right]^{[-2]}\right\}
 =&
 \frac{\xi^k\xi^j}{y^2}
\frac{F(t-r)}r +\\
&P^{jk}
 \left\{\frac{1}y\left[\frac{F(t-r)}r\right]^{[-1]}-
      \int\limits_{-\infty}^{y}
     \frac{1}{\eta^2}
         \int\limits_{-\infty}^{\eta}\frac{F(t^*+\zeta)}\zeta\,
    d\zeta
     d\eta
 \right\}\;,\notag\\
 \label{d32}
 \dksi_{ijk}\left\{\left[\frac{F(t-r)}r\right]^{[-2]}\right\}
 =&
 \frac{1}y
 \left\{
         \left(
        P^{ij}+\frac{\xi^{ij}}{yr}
            \right)\dksi_k
          +P^{jk}\dksi_i
          +\xi^j\dksi_{ik}
 \right\}
\left[\frac{F(t-r)}r\right]^{[-1]}\;,
\end{align}
In the last expression all terms contain at least a first order
derivative of the integral $[F(t-r)/r]^{[-1]}$ with respect to
$\xi^i$. Above it was shown (Eq. \eqref{d11}) that in this case
the integration is eliminated. Therefore the third and higher
order derivatives of $[F(t~-~r)/r]^{[-2]}$ with respect to $\xi^i$
can be expressed in terms of $F(t-r)$ and one does not have to
perform integration. Explicitly evaluating the derivatives in
\eqref{d32} using \eqref{d11} one gets
\begin{align}
\label{d99}
\dksi_{ijk}\left\{\left[\frac{F(t-r)}r\right]^{[-2]}\right\}
 =&\frac{P^{ij}\xi^k+P^{jk}\xi^i+P^{ik}\xi^j}{y^2r}\,F(t-r)\\
&+\frac{\xi^i\xi^j\xi^k}{y^2r^2}\left[\left(\frac{2}{y}-\frac{1}{r}\right)F(t-r)-\dot
F(t-r)\right]\notag\;.
\end{align}
In the equations for trajectory of a photon and gravity-induced
rotation of the polarization plane the integrals
$[F(t-r)/r]^{[-1]}$ and $[F(t-r)/r]^{[-2]}$ are differentiated
with respect to $\tau$ and $\xi^i$ and it can be shown that in all
terms differentiation eliminates integration. In the subsequent
sections we write out solutions to the equations formally in terms
of integrals $[F(t-r)/r]^{[-1]}$ and $[F(t-r)/r]^{[-2]}$, but one
should bear in mind that performing differentiations using the
formulae \eqref{ip1}, \eqref{d11} and \eqref{d32} the integrals
can be expressed in terms of integrands.
\subsection{Solution to the Equations for Trajectory of a Photon}
Using the formulae \eqref{rdif2} and \eqref{rdif3} and rewriting
the equation \eqref{eq1} in new variables one obtains
\begin{equation}
\label{eq2+}
\frac{d^2 x^i(\tau)}{d\tau^2}=
 \ffrac{1}{2}k^\alpha k^\beta\,\dksi_i h_{\alpha\beta}^{\rm can.}-
  \dtau
   \left( k^\alpha h_{i\alpha}^{\rm can.}-
    \ffrac{1}{2}\,k^ik^jk^p q^{\rm can.}_{jp}\right)
    -\dtautau(w^i-k^iw^0)\;,
\end{equation}
where all functions in the right-hand side are taken on the
unperturbed light-ray trajectory (before performing
differentiation). The gauge functions $w^\alpha$ have not yet been
specified which means that equations \eqref{eq2+} are
gauge-invariant. Substituting into \eqref{eq2+} the metric
perturbations expressed in terms of $\tau$ and $\xi^i$ (Eqns.
\eqref{nm}-\eqref{qijs} in Appendix \ref{mnv}) and integrating
once with respect to $\tau$ one obtains \cite{KoKo05}
\begin{equation}
\label{tr1}
         \dot x^i(\tau)=k^i+\dot\Xi^i(\tau,{\bm\xi})\;,
\end{equation}
where
\begin{align}
\label{truk}
         \dot\Xi^i(\tau,{\bm\xi})=&\DXGi+\DXiim+\DXiis\;,
\end{align}
\begin{align}
\label{po23}
\DXGi=&\dtau\left[(\varphi^i-k^i\varphi^0)+(w^i-k^iw^0)\right]\;,\\
\label{xidm}
\DXiim=&2\Mc\left[\dksi_i\frac{1}r\right]^{[-1]}-k^i\frac{2\Mc}r+\\
& 2\dksi_i
    \sum_{l=2}^\infty\sum_{p=0}^l\sum_{q=0}^p
     \frac{(-1)^{l+p-q}}{l!}
       C_l(l-p,p-q,q)H(2-q)
\times\notag
\\
&\left(
    1-\frac{p-q}l
   \right)
       \left(
               1-\frac{p-q}{l-1}
         \right)
            k_{<a_1\hdots a_p}\dksi_{a_{p+1}\hdots a_l>}
    \dtau^q
         \left[\frac{{\Ic}^{(p-q)}_{A_{l}}(t-r)}r\right]^{[-1]}-
\notag    \\
&2\sum_{l=2}^\infty\sum_{p=0}^{l-1}
            \frac{(-1)^{l+p}}{l!}C_l(l-p,p)\left(
    1-\frac{p}l
  \right)
\times\notag\\
&\Biggl\{
    \left(
        1+\frac{p}{l-1}
      \right)
            k_{i<a_1\hdots a_p}\dksi_{a_{p+1}\hdots a_l>}
       \left[\frac{{\Ic}^{(p)}_{A_{l}}(t-r)}r   \right]-
\notag    \\
    &\frac{2p}{l-1}
            k_{<a_1\hdots a_{p-1}}\dksi_{a_p\hdots a_{l-1}>}
        \left[\frac{{\Ic}^{(p)}_{iA_{l-1}}(t-r)}r\right]
             \Biggr\}\;,
  \notag
\end{align}
\begin{align}
\label{xids} \DXiis=
2k_j\epsilon_{jba}\mathcal{S}_b\left[\dksi_{ia}\frac{1}r\right]^{[-1]}
-2\dksi_a\eisbr-&\\
4k_j\dksi_{ia}
    \sum_{l=2}^{\infty}\sum_{p=0}^{l-1}\sum_{q=0}^{p}
        \frac{(-1)^{l+p-q}l}{(l+1)!}
&C_{l-1}\left(l-p-1,p-q,q\right)H(2-q)\times
    \notag
\\
    \left(1-\frac{p-q}{l-1}\right)
&k_{<a_1\hdots a_p}\dksi_{a_{p+1}\hdots a_{l-1}>}
        \dtau^q
        \left[\frac{\epsilon_{jba}{\mathcal {S}}^{(p-q)}_{bA_{l-1}}(t-r)}r\right]^{[-1]}+
    \notag
\\
4\left(\dksi_a-k_a\dt\right) \sum_{l=2}^{\infty}\sum_{p=0}^{l-1}
        \frac{(-1)^{l+p}l}{(l+1)!}
&C_{l-1}\left(l-p-1,p\right)\times
    \notag
\\ \left(1-\frac{p}{l-1}\right)
& k_{<a_1\hdots a_p}\dksi_{a_{p+1}\hdots a_{l-1}>}
        \left[\eislrrp\right]+
    \notag
\\
4k_j \sum_{l=2}^{\infty}\sum_{p=0}^{l-1}
        \frac{(-1)^{l+p}l}{(l+1)!}
C_{l-1}\left(l-p-1,p\right) &k_{<a_1\hdots
a_p}\dksi_{a_{p+1}\hdots a_{l-1}>}
        \left[ \frac {\epsilon_{jba_{l-1}}{\mathcal {S}}^{(p+1)}_{\hat ibA_{l-2}}(t-r)}r\right]\;.
    \notag
\end{align}
Here $\DXGi$ represents the gauge-dependent perturbations,
$\DXiim$ and $\DXiis$ are the perturbations due to the mass and
spin multipole moments, correspondingly. The Heaviside function
$H(p-q)$ is defined by the expression \eqref{H} and
$C_l(l-p,p-q,q)$ are the polynomial coefficients (\ref{pco}). The
gauge functions $\varphi^\alpha$ were introduced as follows: we
collected in the equations \eqref{eq2+} all terms with second and
higher order derivatives with respect to $\tau$ and equated them
to $k^i\varphi^0-\varphi^i$. By introducing the functions
$\varphi^\alpha$ we single out the terms that can be immediately
integrated.  We do not separate $k^i\varphi^0-\varphi^i$ into
$\varphi^0$ and $\varphi^i$, such separation is not unique while
the functions $k^i\varphi^0-\varphi^i$ are uniquely defined (Eqns.
\eqref{phi}-\eqref{phis} in the Appendix \ref{gfa35}).

The gauge functions $w^\alpha$ can be chosen arbitrarily. For the
reasons discussed in Section \ref{gauge} we choose $w^\alpha$
which make our coordinate system ADM-harmonic, that is satisfying
both ADM and harmonic gauge conditions. These functions are given
by the expressions \eqref{nw0} and \eqref{nwi}.

 The second integration of \eqref{eq2+} yields the expressions for trajectory of the
photon
\begin{equation}
\label{tr2} x^i(\tau)=x_N^i+\Delta\Xii\limits_{\scriptscriptstyle
(G)}+\Delta\Xii\limits_{\scriptscriptstyle
(M)}+\Delta\Xii\limits_{\scriptscriptstyle (S)}\;,
\end{equation}
where
\begin{align}
\label{mk4}
\Delta\Xii\limits_{\scriptscriptstyle (G)}\equiv&\XG-\XGo\;,\\
\Delta\Xii\limits_{\scriptscriptstyle (M)}\equiv&\Xiim-\Xiimo\;,\notag\\
\Delta\Xii\limits_{\scriptscriptstyle
(S)}\equiv&\Xiis-\Xiiso\;.\notag
\end{align}
The term
\begin{equation}
\label{pogi} \XG=(\varphi^i-k^i\varphi^0)+(w^i-k^iw^0)\;,
\end{equation}
represents the gauge-dependent part of the trajectory's
perturbation, and the physically meaningful perturbations due to
the mass and spin multipoles are given by
\begin{align}
\label{xim} \Xiim=&2\Mc\left[\dksi_i\frac{1}r\right]^{[-2]}-2\Mc
k^i\left[\frac{1}r\right]^{[-1]}+\\
&2\dksi_i
   \sum_{l=2}^\infty\sum_{p=0}^l\sum_{q=0}^p
     \frac{(-1)^{l+p-q}}{l!}
       C_l(l-p,p-q,q)H(2-q)
\times\notag
\\
&\left(
    1-\frac{p-q}l
   \right)
       \left(
               1-\frac{p-q}{l-1}
         \right)
            k_{<a_1\hdots a_p}\dksi_{a_{p+1}\hdots a_l>}
    \dtau^q
       \left[\frac{{\Ic}^{(p-q)}_{A_{l}}(t-r)}r\right]^{[-2]}-
\notag    \\
&2\sum_{l=2}^\infty\sum_{p=0}^{l-1}
            \frac{(-1)^{l+p}}{l!}
                                  C_l(l-p,p)\left(
    1-\frac{p}l
  \right)\times\notag\\
&\Biggl\{
    \left(
        1+\frac{p}{l-1}
      \right)
            k_{i<a_1\hdots a_p}\dksi_{a_{p+1}\hdots a_l>}
                   \left[\ilrrp\right]^{[-1]}-
\notag    \\
    &\frac{2p}{l-1}
            k_{<a_1\hdots a_{p-1}}\dksi_{a_p\hdots a_{l-1}>}
               \left[\frac{{\Ic}^{(p)}_{iA_{l-1}}(t-r)}r\right]^{[-1]}
             \Biggr\}\;,
  \notag
\end{align}
\begin{align}
\label{xis}
\Xiis=&2k_j\epsilon_{jba}\mathcal{S}_b\left[\dksi_{ia}\frac{1}r\right]^{[-2]}
-2\epsilon_{iba}\mathcal{S}_b\left[\dksi_a\frac{1}r\right]^{[-1]}-
\\
&4k_j\dksi_{ia}
    \sum_{l=2}^{\infty}\sum_{p=0}^{l-1}\sum_{q=0}^{p}
        \frac{(-1)^{l+p-q}l}{(l+1)!}
C_{l-1}\left(l-p-1,p-q,q\right)H(2-q)\times
    \notag
\\
&\left(1-\frac{p-q}{l-1}\right) k_{<a_1\hdots
a_p}\dksi_{a_{p+1}\hdots a_{l-1}>}
        \dtau^q
        \left[\frac{\epsilon_{jba}{\mathcal {S}}^{(p-q)}_{bA_{l-1}}(t-r)}r\right]^{[-2]}+
    \notag
\\
&4\left(\dksi_a-k_a\dt\right) \sum_{l=2}^{\infty}\sum_{p=0}^{l-1}
        \frac{(-1)^{l+p}l}{(l+1)!}
C_{l-1}\left(l-p-1,p\right)\times
    \notag
\\
&\left(1-\frac{p}{l-1}\right)
 k_{<a_1\hdots a_p}\dksi_{a_{p+1}\hdots a_{l-1}>}
        \left[\eislrrp\right]^{[-1]}+
    \notag
\\
&4k_j \sum_{l=2}^{\infty}\sum_{p=0}^{l-1}
        \frac{(-1)^{l+p}l}{(l+1)!}
C_{l-1}\left(l-p-1,p\right)\times
    \notag
\\
&k_{<a_1\hdots a_p}\dksi_{a_{p+1}\hdots a_{l-1}>}
                \left[ \frac {\epsilon_{jba_{l-1}} \mathcal {S}^{(p+1)}_{\hat ibA_{l-2}}(t-r)}r\right]^{[-1]}\;.
    \notag
\end{align}
\subsection{Solution to the Equations for Rotation of the Polarization
Plane}\label{skrot} We rewrite the equation \eqref{skr} in terms
of the variables $\tau$ and $\xi^i$ using the
relationship~\eqref{rdif3} and substitute the metric \eqref{met}
expressed in terms of new variables (Eqns.
\eqref{nm}-\eqref{qijs}):
\begin{equation}
\label{skrnv}
 \frac{d\phi}{d \tau}=\frac{1}2
        k^{\alpha}k^j
            \epsilon_{j\hat{p}\hat q}\dksi_q
                h^{can.}_{\alpha\hat{p}} + \frac{1}2k^j
\epsilon_{j\hat p \hat q}
        \dksi_{q\tau}
        w^{\hat p}\;.
\end{equation}
Integrating with respect to $\tau$ and changing the order of
differentiation and integration yields
\begin{equation}
\label{dfi} \phi=\fig+\fim+\fis+\phi_0\;,
\end{equation}
where $\phi_0$ is the constant angle characterizing the initial
orientation of the polarization ellipse in the plane formed by the
vectors ${\bm e}_{(1)}$ and ${\bm e}_{(2)}$. The terms $\fig$,
$\fim$ and $\fis$ describe the contributions due to
gauge-dependent terms, terms depending on mass- and spin-type
moments correspondingly.

The gauge-dependent part of the Skrotskii effect can be obtained
immediately, since the gauge dependent terms appear in the
equation \eqref{skrnv} under the derivative with respect to $\tau$
(cf. Eq. \eqref{ip1}). The integration yields
\begin{equation}
\label{sf79} \fig=\frac{1}2k^j \epsilon_{j\hat p\hat q}
        \dksi_{q}\left(w^{\hat p}+\chi^{\hat p}\right)\;,
\end{equation}
where the gauge functions $w^i$ and $\chi^i$ are given by the
equations \eqref{wi} and \eqref{cccp}. The gauge functions
$\chi^i$ were introduced by collecting in the equation
\eqref{skrnv} all terms that can be eliminated by a gauge
transformation.

 The Skrotskii effect due to the mass and spin
multipoles of the isolated system is given by
\begin{align}
\label{fim}
 \fim(\tau)=&
2\sum_{l=2}^\infty\sum_{p=0}^l
     \frac{(-1)^{l+p}}{l!}
     C_l(l-p,p)
\frac{l-p}{l-1}\\
   &k_{<a_1\hdots a_p}\dksi_{a_{p+1}\hdots a_l>j}
         \left[\frac{
            \epsilon_{j\hat{b}a_l}\Ic^{(p)}_{\hat{b}A_{l-1}}(t-r)}r
       \right]^{[-1]}\;,\notag
\end{align}
and
\begin{align}
\label{fis} \fis(\tau) =& 2\sum_{l=1}^\infty\sum_{p=0}^l
    \frac{(-1)^{l+p}l}{(l+1)!}
C_l(l-p,p) \left ( 1-\frac {p}l\right )\\
    &\left [1+H(l-1)
        \left (1-\frac{2p}{l-1}\right )
    \right ]
    k_{<a_1\hdots a_p}\dksi_{a_{p+1}\hdots a_l>}
         \left[
\frac{\Sc^{(p+1)}_{A_{l}}(t-r)}r\right]^{[-1]}\;.\notag
\end{align}
Just like in the case with the equations for light-ray path
\eqref{xidm}-\eqref{xids} and \eqref{xim}-\eqref{xis}
differentiation eliminates integrals in the right-hand sides of
the equations \eqref{fim}-\eqref{fis} as it was shown in Section
\ref{methsol}.
\section{Observable Relativistic Effects}
In this section we give the expressions for the observable
relativistic effects of time delay and gravitational deflection of
light. The expressions for another observable effect -- rotation
of the polarization plane were essentially given in Section
\ref{skrot}.
\subsection{Time Delay}
In the paper~\cite{KSGU99} the following expression was obtained
for the time of propagation of the electromagnetic signals through
the gravitational field of an isolated source from the emitter to
the observer
\begin{align}
\label{td} t-t_0=&
         |{\bm x}-{\bm x}_0|+\Delta(\tau,\tau_0)\;,\\
         \label{t93}
         \Delta(\tau,\tau_0)=&\DelG+\Delm+\Dels\;,\\
\intertext{where $x^\alpha_0=(t_0,{\bm x}_0)$ and
$x^\alpha=(t,{\bm x})$ are the coordinates of the points of
emission and observation of the signal, $\DelG$, $\Delm$ and
$\Dels$ are functions describing the delay of the electromagnetic
signal due to the gauge-dependent terms, mass and spin multipoles
of the gravitational field of the isolated system correspondingly.
These functions are expressed  in terms of the perturbations of
the signal's trajectory as}
\label{t78} \DelG=&-k_i\left[\XG
                -\XGo
             \right],\\
\label{tdm} \Delm=&
     -k_i\left[\Xiim
                -\Xiimo
             \right],\\
\label{tds} \Dels=&
     -k_i\left[\Xiis
                -\Xiiso
             \right]\;.
\end{align}

We note that the expressions \ref{t78} give the effect with
respect to the coordinate time which has to be converted to the
time measured by the observer. It is shown in \cite{KSGU99} that
if the observer's velocity is negligible with respect to the
global ADM-harmonic coordinate system the relationship between the
coordinate time and proper time $T$ of the observer is given by
\begin{eqnarray}
\label{tra} T&=&\left(1-\frac{{\cal
M}}{r}\right)\left(t-t_i\right)\;,
\end{eqnarray}
where $t_{\rm i}$ is the initial epoch of observation. In most
cases the distance $r$ is much grater than ${\cal M}$ and
coordinate time coincide with the proper time of the observer. If
observer is moving with respect to the global coordinate system,
an additional Lorentz transformation to the rest frame of the
observer has to be performed. This case was considered in
\cite{KoKo05}.
\subsection{Deflection of Light}
The expression for observable vector towards the source of light
calculated with respect to the observer's reference frame was
derived in ~\cite{KSGU99} and reads
\begin{equation}
\label{s}
   s^i\left(\tau,{\bm \xi}\right)=K^i+
            \alpha^i\left(\tau,{\bm \xi}\right)+
            \beta^i\left(\tau,{\bm \xi}\right)-
            \beta^i\left(\tau_0,{\bm \xi}\right)+
            \gamma^i\left(\tau,{\bm \xi}\right),
\end{equation}
where
\begin{align}
\label{K}
    K^i=&\frac {x^i-x^i_0}{|{\bf x}-{\bf x}_0|}
\\
\intertext{ is the unit vector in the direction "observer --
source of light", defined in the global coordinate system;}
\label{a}
     \alpha^i\left(\tau,{\bm \xi}\right)=
    &-P^i_j\DXij
           \\
\intertext{is the vector describing the angle of light
deflection;} \label{b} \beta^i\left(\tau,{\bm \xi}\right)=
    &\frac{P^i_j
        \left[\Xijm
            +\Xijs
            \right]}{|{\bf x}-{\bf x}_0|}
\\
\intertext{is relativistic corrections, introduced by the
relationship}
 k^i=
    &-K^i- \beta^i\left(\tau,{\bm \xi}\right)+
\beta^i\left(\tau_0,{\bm \xi}\right).
\\
\intertext{The term} \label{g}
     \gamma^i\left(\tau,{\bm \xi}\right)=
    &-\ffrac{1}2P^{ij}k^qh_{jq}^{TT}(t,{\bf x})
\end{align}
appears as a result of transformation from the the global
ADM-harmonic system to the local frame of the observer and
describes the perturbations of the observer's coordinate system
due to the gravitational waves emitted by the isolated system. It
was assumed that the observer is at rest with respect to the
global ADM-harmonic system. In the case when the observer is
moving with respect to the global system one has to perform an
additional Lorentz transformation to the rest frame of the
observer.
\appendix
\section{The Metric Tensor in Terms of Variables $\bm\xi$ and $\tau$}\label{mnv} 
The expressions for perturbations of the metric
tensor \eqref{bm} -- \eqref{em} in variables $\bm\xi$ and $\tau$
have the form as follows
\begin{align}
\label{nm}
 h_{00}^{\rm can.}=&\frac {2\Mc}r
                        +2
                        \sum_{l=2}^\infty\sum_{p=0}^l\sum_{q=0}^p
                        \h\limits_{{\scriptscriptstyle (M)}}{}\!\!_{00}^{lpq}(t^*,\tau,{\bm\xi})\;,
\\
    h_{0i}^{\rm can.}=&-\frac{2\epsilon_{iba}\Sc^bN^a}{r^2}
                    +4\sum_{l=2}^\infty\sum_{p=0}^{l-1}\sum_{q=0}^p\left[
                                   \h\limits_{{\scriptscriptstyle (S)}}{}\!\!_{0i}^{lpq}(t^*,\tau,{\bm\xi})
                    +\h\limits_{{\scriptscriptstyle (M)}}{}\!\!_{0i}^{lpq}(t^*,\tau,{\bm\xi})\right]\;,
\\
\label{nhij}
h_{ij}^{\rm can.}=&\delta_{ij}h_{00}^{\rm can.}+q_{ij}^{\rm can.}\;,
\\
\label{nqij}
q_{ij}^{\rm can.}=&4\sum_{l=2}^\infty\sum_{p=0}^{l-2}\sum_{q=0}^p
                                     \q\limits_{{\scriptscriptstyle (M)}}{}\!\!_{ij}^{lpq}(t^*,\tau,{\bm\xi})
                   -8\sum_{l=2}^\infty\sum_{p=0}^{l-2}\sum_{q=0}^p
                                 \q\limits_{{\scriptscriptstyle
                                 (S)}}{}\!\!_{ij}^{lpq}(t^*,\tau,{\bm\xi})\;,
\end{align}
where
\begin{align}
 \h\limits_{{\scriptscriptstyle (M)}}{}\!\!_{00}^{lpq}(t^*,\tau,{\bm\xi})=&
\label{hoom}
        \frac{(-1)^{l+p-q}}{l!}C_l(l-p,p-q,q)\\
&k_{<a_1\ldots a_p}\dksi_{a_{p+1}\ldots a_l>}
                  \dtau^{q}
                     \left[\frac{\Ic^{(p-q)}_{A_{l}}(t-r)}r\right]\;,\notag
                     \\
\label{hoim}
 \h\limits_{{\scriptscriptstyle(M)}}{}\!\!_{0i}^{lpq}(t^*,\tau,{\bm\xi})=&
        \frac{(-1)^{l+p-q}}{l!}C_{l-1}(l-p-1,p-q,q)\\
 &k_{<a_1\ldots a_p}\dksi_{a_{p+1}\ldots a_{l-1}>}
                              \dtau^{q}
                              \left[\frac{\Ic^{(p-q+1)}_{iA_{l-1}}(t-r)}r\right],\notag
                                             \\
\label{hois}
\h\limits_{{\scriptscriptstyle(S)}}{}\!\!_{0i}^{lpq}(t^*,\tau,{\bm\xi})=&
  \frac{(-1)^{l+p-q}\;l}{(l+1)!}C_{l-1}(l-p-1,p-q,q)\\
&(\dksi_a+k_a\dtau-k_a\dt)
      k_{<a_1\ldots a_p}\dksi_{a_{p+1}\ldots a_{l-1}>}
                              \dtau^{q}
                            \left[\frac{\epsilon_{iab}\Sc^{(p-q)}_{bA_{l-1}}(t-r)}r   \right]\;,\notag
       \\
\label{qijm}
 \q\limits_{{\scriptscriptstyle(M)}}{}\!\!_{ij}^{lpq}(t^*,\tau,{\bm\xi})=&
\frac{(-1)^{l+p-q}}{l!}C_{l-2}(l-p-2,p-q,q)\\
      &k_{<a_1\ldots a_p}\dksi_{a_{p+1}\ldots a_{l-2}>}
                              \dtau^{q}
                              \left[\frac{\Ic^{(p-q+2)}_{ijA_{l-2}}(t-r)}r\right]\;,\notag
         \\
\label{qijs}
\q\limits_{{\scriptscriptstyle(S)}}{}\!\!_{ij}^{lpq}(t^*,\tau,{\bm\xi})=&
        \frac{(-1)^{l+p-q}l}{(l+1)!}C_{l-2}(l-p-2,p-q,q)\\
        &(\dksi_a+k_a\dtau-k_a\dt)
      k_{<a_1\ldots a_p}\dksi_{a_{p+1}\ldots a_{l-2}>}
                              \dtau^{q}
                              \left[\frac{\epsilon_{ba(i}\Sc^{(p-q+1)}_{j)bA_{l-2}}(t-r)}r\right]\;.\notag
\end{align}
All quantities in the right side of expressions
\eqref{hoom}--\eqref{qijs}, which are explicitly shown as
functions of $x^i$, $r=|{\bm x}|$ and $t$, must be understood as
taken on the unperturbed light-ray trajectory and expressed in
terms of $\xi^i$, $d=|{\bm\xi}|$, $\tau$ and $t^*$ in accordance
with the equations \eqref{tt*}, \eqref{rtd}. For example, the
ratio $\Ic^{(p-q)}_{A_{l}}(t-r)/r$ in equation \eqref{hoom} must
be understood as
\begin{equation}
\label{kob}
\frac{\Ic^{(p-q)}_{A_{l}}(t-r)}r\equiv\frac{\Ic^{(p-q)}_{A_{l}}(t^*+\tau-\sqrt{\tau^2+d^2})}{\sqrt{\tau^2+d^2}}\;,
\end{equation}
and the same replacement rule is applied to the other equations.

\section{Gauge Functions}\label{gfa35}
 Gauge functions ${w}^\alpha$, generating the
coordinate transformation from the canonical harmonic coordinate
system to the ADM-harmonic, are given by equations \eqref{w0},
\eqref{wi}. They transform the metric tensor as follows
\begin{equation}
\label{amq}
h_{\alpha\beta}^{\rm can.}=h_{\alpha\beta}-\partial_\alpha w_\beta-\partial_\beta w_\alpha\;,
\end{equation}
where $h_{\alpha\beta}^{\rm can.}$ is the canonical form of the
metric tensor in  harmonic coordinates given by equations
\eqref{bm}--\eqref{em} and $h_{\alpha\beta}$ is the metric tensor
given in the ADM-harmonic coordinates by equations
\eqref{adm1}--\eqref{adm4}.

The gauge functions taken on the light-ray trajectory and
expressed in terms of the variables $\xi^i$ and $\tau$ can be
written down in the form
\begin{eqnarray}
\label{nw0}
{w}^0&=&\sum_{l=2}^\infty\sum_{p=0}^l\sum_{q=0}^p\int^{\tau+t^*}_{-\infty}du
    \h\limits_{{\scriptscriptstyle (M)}}{}\!\!_{00}^{lpq}(u,\tau,{\bm\xi}),
\\
\label{nwi}
{w}^i&=&(\dksi_i+k_i\dtau-k_i\dt)
    \sum_{l=2}^\infty\sum_{p=0}^l\sum_{q=0}^p
   \int^{\tau+t^*}_{-\infty}dv\int^{v}_{-\infty}du
        \h\limits_{{\scriptscriptstyle (M)}}{}\!\!_{00}^{lpq}(u,\tau,{\bm\xi})\\\nonumber&&
        -4\sum_{l=2}^\infty\sum_{p=0}^{l-1}\sum_{q=0}^p
                    \int^{\tau+t^*}_{-\infty}du\left[
                    \h\limits_{{\scriptscriptstyle
                     (M)}}{}\!\!_{0i}^{lpq}(u,\tau,{\bm\xi})
                       +\h\limits_{{\scriptscriptstyle (S)}}{}\!\!_{0i}^{lpq}(u,\tau,{\bm\xi})\right]\;,
\end{eqnarray}
where $ \h\limits_{{\scriptscriptstyle (M)}}{}\!\!_{00}^{lpq}(u,\tau,{\bm\xi}),
\h\limits_{{\scriptscriptstyle (M)}}{}\!\!_{0i}^{lpq}(u,\tau,{\bm\xi})$ and
  $\h\limits_{{\scriptscriptstyle (S)}}{}\!\!_{0i}^{lpq}(u,\tau,{\bm\xi})$
are defined by the Eqns. \eqref{hoom}, \eqref{hoim} and
\eqref{hois} after making use of the substitution $t^*\rightarrow
u$.

Linear combination $k^i\varphi^0-\varphi^i$ of the gauge-dependent
functions $\varphi^\alpha$ introduced in equation \eqref{po23} is
given by the expressions
\begin{equation}
\label {phi}
 k^i\varphi^0-\varphi^i=(k^i\phim^0-\phim^i)+(k^i\phis^0-\phis^i)\;,
\end{equation}
\begin{align}
\label{phim}k^i\phim^0-\phim^i=
2\dksi_i\sum_{l=2}^\infty\sum_{p=2}^l\sum_{q=2}^p
          \frac{(-1)^{l+p-q}}{l!}
&                                C_l(l-p,p-q,q)\times
       \\
            \left(
    1-\frac{p-q}l
   \right)
       \left(
               1-\frac{p-q}{l-1}
         \right)
&
           k_{<a_1\ldots a_p}\dksi_{a_{p+1}\ldots a_l>}
                 \dtau^{q-2}
         \left[\frac{\Ic^{(p-q)}_{A_l}(t-r)}{r}\right]+
\notag \\
     2\sum_{l=2}^\infty\sum_{p=1}^l\sum_{q=1}^p
           \frac{(-1)^{l+p-q}}{l!}C_l(l
&                                      -p,p-q,q)\times
\notag \\
      \left(1-\frac{p-q}l\right)\left\{
           \left(1+\frac{p-q}{l-1}\right)
         \right.
&
           k_{i<a_1\ldots a_p}\dksi_{a_{p+1}\ldots a_l>}
                  \dtau^{q-1}
         \left[\frac{\Ic^{(p-q)}_{A_l}(t-r)}{r}\right]-
\notag \\
       2\frac{p-q}{l-1}
&
          \left.
           k_{<a_1\ldots a_{p-1}}\dksi_{a_p\ldots a_{l-1} >}
                   \dtau^{q-1}
            \left[\frac{\Ic^{(p-q)}_{iA_{l-1}}(t-r)}r\right]
          \right\},
\notag
\end{align}
\begin{align}
\label{phis}k^i\phis^0-\phis^i=&
2\frac{\epsilon_{iab}k^a\Sc^b}{r}+\\&4k_j\dksi_{ia}
    \sum_{l=3}^{\infty}\sum_{p=2}^{l-1}\sum_{q=2}^{p}
        \frac{(-1)^{l+p-q}\;l}{(l+1)!}
C_{l-1}\left(l-p-1,p-q,q\right)\times\notag
\\
    &\left(1-\frac{p-q}{l-1}\right)
k_{<a_1\hdots a_p}\dksi_{a_{p+1}\hdots a_{l-1}>}
        \dtau^{q-2}\left[\frac{\epsilon_{jab}\Sc^{(p-q)}_{bA_{l-1}}(t-r)}{r}\right]-
    \notag
\\
&4\left(\dksi_a-k_a\dt\right)
\sum_{l=2}^{\infty}\sum_{p=1}^{l-1}\sum_{q=1}^{p}
        \frac{(-1)^{l+p-q}l}{(l+1)!}
C_{l-1}\left(l-p-1,p-q,q\right)\times
    \notag
\\& \left(1-\frac{p-q}{l-1}\right)
 k_{<a_1\hdots a_p}\dksi_{a_{p+1}\hdots a_{l-1}>}
        \dtau^{q-1}\left[\frac{\epsilon_{iab}\Sc^{(p-q)}_{bA_{l-1}}(t-r)}{r}\right]-
    \notag
\\&4k_a
\sum_{l=2}^{\infty}\sum_{p=0}^{l-1}\sum_{q=0}^{p}
        \frac{(-1)^{l+p-q}l}{(l+1)!}
C_{l-1}\left(l-p-1,p-q,q\right)\times
    \notag
\\& \left(1-\frac{p-q}{l-1}\right)
k_{<a_1\hdots a_p}\dksi_{a_{p+1}\hdots a_{l-1}>}
        \dtau^q\left[\frac{\epsilon_{iab}\Sc^{(p-q)}_{bA_{l-1}}(t-r)}{r}\right]+
    \notag
\\&
4k_j
\sum_{l=2}^{\infty}\sum_{p=1}^{l-1}\sum_{q=1}^{p}
        \frac{(-1)^{l+p-q}l}{(l+1)!}
C_{l-1}\left(l-p-1,p-q,q\right)\times
    \notag
\\
&k_{<a_1\hdots a_p}\dksi_{a_{p+1}\hdots a_{l-1}>}
        \dtau^{q-1}\left[ \frac {\epsilon_{jba_{l-1}}\mathcal {S}^{(p-q+1)}_{\hat ibA_{l-2}}(t-r)}r\right]
    \notag
\end{align}
Gauge-dependent term generated by equation \eqref{sf79} for the
rotation of the plane of polarization of electromagnetic wave is a
pure spatial vector $\chi^i$ that can be decomposed in two linear
parts corresponding to the mass and spin multipoles:
\begin{equation}
\label{cccp}
\chi^i=\chim^i+\chis^i\;,
\end{equation}
where
\begin{align}
\label{chim} \chim^i=
 &4\sum_{l=2}^\infty\sum_{p=1}^{l-1}\sum_{q=1}^p
      \frac{(-1)^{l+p-q}}{l!}
      C_{l-1}(l-p-1,p-q,q)
 \left(1-\frac{p-q}{l-1}
            \right)
\\
 & \phantom{oooooooooo}\times
    k_{<a_1\hdots a_p}\dksi_{a_{p+1}\hdots a_{l-1}>}
     \dtau^{q-1}
      \left[\frac{
      \Ic^{(p-q+1)}_{iA_{l-1}}(t-r)
                }r\right]\;,
\notag    \\
 \label{chis}
  \chis^i=
 &-4\sum_{l=1}^\infty\sum_{p=0}^{l-1}\sum_{q=0}^p
    \frac{(-1)^{l+p-q}l}{(l+1)!}
 C_{l-1}(l-p-1,p-q,q)
 \left [ 1-
        \frac{p-q}{l-1}\,H(l-1)
    \right ]
\\
 &
 \times
 \left[
 H(q)(\dksi_a-k_a\dt)+k_a\dtau
 \right]
 k_{<a_1\hdots a_p}\dksi_{a_{p+1}\hdots a_{l-1}>}
 \dtau^{q-1}\left[
 \frac{\epsilon_{iba}\Sc^{(p-q)}_{bA_{l-1}}(t-r)}r\right]
 \notag \\
 &+4\sum_{l=3}^\infty\sum_{p=1}^{l-1}\sum_{q=1}^p
    \frac{(-1)^{l+p-q}l}{(l+1)!}
 C_{l-1}(l-p-1,p-q,q)
 \left ( 1-\frac {p}l\right )
    \left (1-\frac{q}p
    \right )
 \notag  \\
 &\phantom{ooooooooo}\times
    k_{<a_1\hdots a_p}\dksi_{a_{p+1}\hdots a_{l-1}>a}
 \dtau^{q-1}\left[
 \frac{\epsilon_{baa_{l-1}}\Sc^{(p-q)}_{ibA_{l-2}}(t-r)}r\right]\;.
  \notag
\end{align}
\pagebreak

\end{document}